\documentclass[referee,pdflatex,sn-nature]{sn-jnl}%

\usepackage{graphicx}%
\usepackage{multirow}%
\usepackage{amsmath,amssymb,amsfonts}%
\usepackage{amsthm}%
\usepackage{mathrsfs}%
\usepackage[title]{appendix}%
\usepackage{xcolor}%
\usepackage{textcomp}%
\usepackage{manyfoot}%
\usepackage{booktabs}%
\usepackage{algorithm}%
\usepackage{algorithmicx}%
\usepackage{algpseudocode}%
\usepackage{listings}%
\usepackage{epstopdf}

\usepackage[version=4]{mhchem}
\usepackage[normalem]{ulem}

\begin{document}

\title[Article Title]
{
Near-surface Defects Break Symmetry in Water Adsorption on CeO$_{2-x}$(111) 
}

\author*[1]{\fnm{Oscar} \sur{Custance}}
\email{custance.oscar@nims.go.jp}
\equalcont{These authors contributed equally to this work.}

\author[2]{\fnm{Manuel} \sur{Gonz\'{a}lez-Lastre}}
\equalcont{These authors contributed equally to this work.}

\author[3]{\fnm{Kyungmin} \sur{Kim}}

\author[2,4]{\fnm{Estefan\'{i}a} \sur{Fernandez-Villanueva}}

\author[2,5]{\fnm{Pablo} \sur{Pou}}

\author[3]{\fnm{Masayuki} \sur{Abe}}

\author[1]{\fnm{Hossein} \sur{Sepehri-Amin}}

\author[1]{\fnm{Shigeki} \sur{Kawai}}

\author[4]{\fnm{M. Ver\'{o}nica} \sur{Ganduglia-Pirovano}}

\author*[2,5]{\fnm{Ruben} \sur{Perez}}
\email{ruben.perez@uam.es}

\affil [1]{\orgname{National Institute for Materials Science (NIMS)}, \orgaddress{\street{1-2-1 Sengen}, \city{Tsukuba}, \postcode{305-0047}, \state{Ibaraki}, \country{Japan}}}

\affil [2]{\orgdiv{Departamento de F\'{i}sica Te\'{o}rica de la Materia Condensada}, \orgname{Universidad Aut\'{o}noma de Madrid}, \orgaddress{ \postcode{28049}, \state{Madrid}, \country{Spain}}}

\affil[3]{\orgdiv{Graduate School of Engineering Science}, \orgname{Osaka University}, \orgaddress{\street{1-3 Machikaneyama}, \city{Toyonaka}, \postcode{560-0043}, \state{Osaka}, \country{Japan}}}

\affil [4]{\orgdiv{Instituto de Cat\'{a}lisis y Petroleoqu\'{i}mica (CSIC)}, \orgaddress{ \postcode{28049}, \state{Madrid}, \country{Spain}}}

\affil [5]{\orgdiv{Condensed Matter Physics Center (IFIMAC)}, \orgname{Universidad Aut\'{o}noma de Madrid}, %
\orgaddress{ \postcode{28049}, \state{Madrid}, \country{Spain}}}

\abstract{
Water interactions with oxygen-deficient cerium dioxide (CeO$_2$) surfaces are central to hydrogen production and catalytic redox reactions, but the atomic-scale details of how defects influence adsorption and reactivity remain elusive. Here, we unveil how water adsorbs on partially reduced CeO$_{2-x}$(111) using atomic force microscopy (AFM) with chemically sensitive, oxygen-terminated probes, combined with first-principles calculations. Our AFM imaging reveals water molecules as sharp, asymmetric boomerang-like features radically departing from the symmetric triangular motifs previously attributed to molecular water. Strikingly, these features localize near subsurface defects. While the experiments are carried out at cryogenic temperature, water was dosed at room temperature, capturing configurations relevant to initial adsorption events in catalytic processes. Density functional theory identifies Ce$^{3+}$ sites adjacent to subsurface vacancies as the thermodynamically favored adsorption sites, where defect-induced symmetry breaking governs water orientation. Force spectroscopy and simulations further distinguish Ce$^{3+}$ from Ce$^{4+}$ centers through their unique interaction signatures. By resolving how subsurface defects control water adsorption at the atomic scale, this work demonstrates the power of chemically selective AFM for probing site-specific reactivity in oxide catalysts, laying the groundwork for direct investigations of complex systems such as single-atom catalysts, metal–support interfaces, and defect-engineered oxides.
\\
}

\maketitle

Cerium dioxide (CeO$_2$, or ceria) is a cornerstone material in catalysis, widely used both as an active component and as a support, owing to its exceptional ability to store, release, and transport oxygen\cite{Trovarelli2013,Fornasiero2016}. This redox flexibility is rooted in the ease of reversible Ce$^{4+}$/Ce$^{3+}$ reduction, which enables ceria to form and heal oxygen vacancies under operating conditions. As a result, ceria exhibits remarkable versatility across a broad range of applications, from automobile exhaust gas treatment~\cite{Dresselhaus-2001-414,Gorte2010} and hydrogen production~\cite{Deluga-2004-993} to solid oxide fuel cells~\cite{Park-2000-265} and emerging biomedical technologies~\cite{XuNPG2014,Pulido-ReyesSciRep2015}. In common with other reducible oxides~\cite{titania1,titania2}, the type, density, and spatial distribution of these vacancies profoundly influence catalytic performance. Among the molecular species central to redox catalysis, water plays a pivotal role, particularly in reactions such as the water–gas shift~\cite{Rodriguez-2007-1757} and thermochemical splitting cycles~\cite{Bhosale2019}. Understanding how water interacts with defective ceria surfaces at the atomic scale is therefore essential for the rational design of next-generation catalysts and redox-active materials.

Atomic force microscopy (AFM) has proven to be a powerful tool to explore metal oxide surfaces at the atomic scale, as well as adsorbates at them~\cite{Stetsovych2015IndentificationAnataseAFM, SetvinKTaO001Science2018, DieboldSetvin_ForceSpectroscopy, DiebolCoonKTaOScience, DiebolAlimunaScience2024}.
Constant height imaging with functionalized probes~\cite{Gross2009ChemicalStructureOfaMolecule, Moenig_ASCNano2016_WithRuben, Moenig_ASCNano2024_Standarization} has empowered this technique with unprecedented resolution that has been deployed, among other feats, to study individual water molecules and properties of water networks on several surface systems~\cite{ShiotariNatComm-AFMH20, JiangYingInterfacialWaterWithAFM, JiangYingEigenZundelCations, JiangYingSuperlubricityWaterBNwithAFM, JiangYingPremeltingIce}.

Molecular water on ceria surfaces has been studied with AFM~\cite{Gritschneder_2005, Gritschneder_Water_2007}, as well as with scanning tunneling microscopy (STM)~\cite{Thornton2010_Water}.
For both techniques, and for sample temperatures ranging from room temperature to 10~K, water molecules on the CeO$_2$(111) surface were imaged as wide triangular features extending over three surface oxygen atoms.
At variance with other oxides,
in this ceria surface, molecular water can exist in two configurations \textemdash molecular form and hydroxyl pair\textemdash\space that are almost degenerate in energy~\cite{note0, DeliaWaterOnCeria111, Rockert2020, Zou2023}. 
In the molecular state, water adsorbs with the oxygen atom on a Ce$^{4+}$ site, slightly shifted toward a neighboring surface oxygen, which forms a bond with one of the hydrogen atoms while the second hydrogen points upwards and slightly to one side~\cite{DeliaWaterOnCeria111}. 
The proton involved in the hydrogen bond can hop between the water molecule and the surface oxygen, forming a hydroxyl pair that is stabilized by a hydrogen bond between the oxygen atop the cerium site and the transferred proton of similar length and strength to the one in the molecular case.
Both configurations are connected through a transition state with an energy barrier below 0.10~eV~\cite{DeliaWaterOnCeria111}.
By symmetry, there are six equivalent adsorption configurations, which are also connected by low energy barriers; $\sim 0.01~eV$ to 0.14~eV depending on the atomic path involved in the transition~\cite{DeliaWaterOnCeria111}.

In this work, we study the adsorption of molecular water on a partially reduced CeO$_{2-x}$(111) surface using AFM operated at cryogenic temperatures (T$\sim$4.8~K) with copper-oxide functionalized probes~\cite{Moenig_ASCNano2016_WithRuben, Moenig_ASCNano2024_Standarization}.
Compared with the wide triangular features previously reported,
the strikingly higher resolution achieved in this work provides images
of water molecules as sharp asymmetric boomerangs, connecting the cerium adsorption site with two of the neighboring oxygen atoms; a result that challenges our interpretation of water adsorption on the CeO$_2$(111) surface.
To clarify these new findings, we combine constant-height AFM imaging and force spectroscopy~\cite{Lantz01-Science-ForceSpectroscopy, Sugimoto07-Nature-ChemicalIdentification} with Density Functional Theory (DFT) simulations. 
The use of
the water molecules as local markers~\cite{Stetsovych2015IndentificationAnataseAFM} 
enables us to assign atomic specificity to the 
frequency shift ($\Delta f$) curves (see Methods) measured over the 
atomic sites of the CeO$_{2}$(111) surface, and address the features observed in the constant-height AFM images.
By comparing these 
$\Delta f$ curves
with theoretical predictions, we identified a model 
probe that reproduces well the experiments. This model is represented by a rigid carbon monoxide (CO) molecule that mimics the oxygen-terminated 
probe apex in the experiments, plus a long-range component that accounts for the van der Waals (vdW) interaction with the mesoscopic part of the probe.

We build on this 
model
to understand 
experimental $\Delta f$ curves sampling the atomic sites visited by the water molecule.
Our analysis highlights the key role played by 
the response of the water molecule to the forces exerted by the probe during the experiments
to explain the observed sharp features in the AFM images, and suggests the particular adsorption environment created by a Ce$^{3+}$ near a vacancy as the explanation for the asymmetry.
Furthermore, simulated 
$\Delta f$ curves
predict a stronger attractive interaction on the Ce$^{3+}$ sites relative to the Ce$^{4+}$. 
This different interaction 
with the probe 
originates from
distortions on the topmost oxygen layer caused by the excess charge localized on Ce$^{3+}$, 
and it
paves the way for a possible experimental identification of these so far elusive defects.

Our work reveals the potential of AFM 
as a powerful atomic-resolution local technique to address and understand problems involving oxide nanostructures and their role in catalysis, such as the study of single-atom catalysts 
due to the possibility of accessing the local reactivity.

\section*{Results and discussion}\label{Results}

\begin{figure}[t!]
\centering
\includegraphics[width=0.9\textwidth]{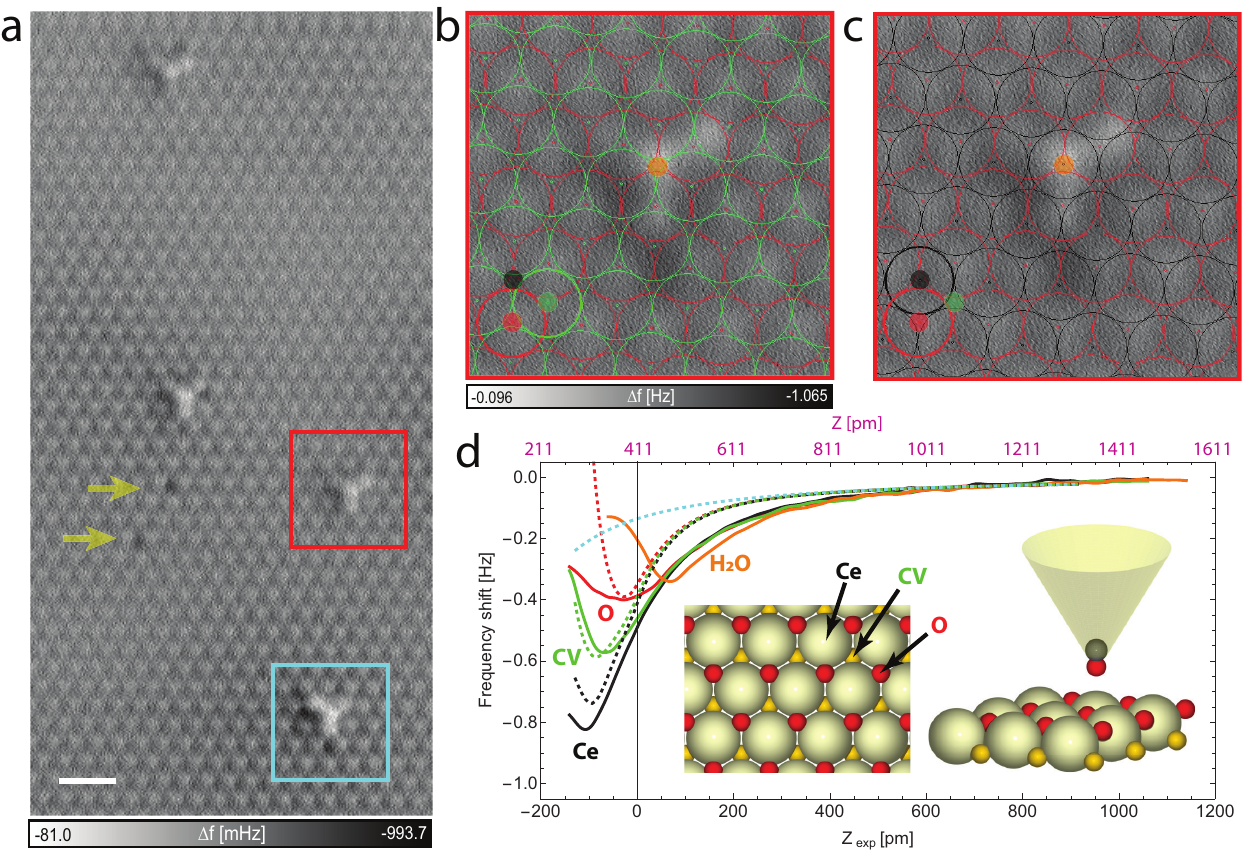}
\caption{{\bf Atom identification using water molecules as atomic markers
}
{\bf a}, Atomic-resolution constant height AFM image of a CeO$_{2}$(111) surface area displaying four water molecules, two of them enclosed by a red and a cyan square.
{\bf b} and {\bf c}, Water molecule highlighted with a red square in {\bf a} with hexagonal lattices in red, green and black superimposed over the atomic sites of the CeO$_{2}$(111) surface: cerium (Ce), oxygen (O) and coordination vacancy (CV).
The comparison of pairs of lattices with the adsorption position of the water molecule enables us to assign the black, red and green lattices to Ce, O and CV, respectively.
{\bf d}, 
Frequency shift ($\Delta f$) curves (see Methods)
measured at the top of the water molecule (orange), and on an O (red), a Ce (black) and a CV (green) site far from the molecule in ({\bf b} and {\bf c}).
The dotted curves are calculated $\Delta f$ curves from DFT interatomic forces (see Methods) obtained over the three CeO$_{2}$(111) surface sites using a probe modeled by a rigid CO molecule (inset).
A long range van der Waals interaction derived from fitting the experimental curves at large distances (cyan dotted line) has been added to each of the calculated curves to include the interaction with the mesoscopic part of the probe. 
The origin in the experimental distance axis ($Z_{exp}$) corresponds to the probe-surface separation the images in ({\bf b}) and ({\bf c}) were acquired.
Comparing with the calculated $\Delta f$ curves, this separation is equivalent to $\sim$ 411~pm with respect to the plane defined by the centres of the top-most surface oxygen atoms. 
 Acquisition parameters were: an oscillation amplitude ($A$) of 60~pm, and a free oscillation resonant frequency ($f_0$) of 994230 Hz.
 Yellow arrows in {\bf a} point to Ce atoms with an unusual contrast.
 The scale bar corresponds to a 1~nm distance.
}
\label{Fig1}
\end{figure}

Figure~\ref{Fig1}a shows an atomic resolution AFM image of an area of the CeO$_{2}$(111) surface that presents four water molecules (see Fig.~S1 for a general view of the surface region).
Since it is well established that molecular water adsorbs on top of a cerium atom~\cite{Gritschneder_Water_2007,DeliaWaterOnCeria111}, these water molecules were used as
markers~\cite{Stetsovych2015IndentificationAnataseAFM} to assign chemical specificity to the atomic features observed in the AFM images and the 
$\Delta f$ curves~\cite{Lantz01-Science-ForceSpectroscopy, Sugimoto07-Nature-ChemicalIdentification} 
measured over them (see the caption of Fig~\ref{Fig1} for details).

The analysis of these $\Delta f$ curves is paramount for the correct interpretation of atomic resolution in constant height AFM images. 
The cerium atoms present
the most attractive interaction with the probe, the oxygen atoms show the weakest, 
and the coordination vacancy
displays an intermediate behavior, which can be attributed to the sum of forces exerted on the probe by the surface atoms surrounding it (Fig.~\ref{Fig1}d).
The behavior of the experimental curves over the cerium and oxygen sites suggests an oxygen-terminated apex,
which is consistent with the 
probe conditioning at
the copper oxide areas~\cite{Moenig_ASCNano2016_WithRuben} 
coexisting with the ceria islands (see Methods).
The nature of the probe apex
is further confirmed by theoretical calculations of the 
$\Delta f$ curves
using a model probe that combines the short-range contribution from DFT-calculated forces with the long-range probe-surface interaction (inset in Fig.~\ref{Fig1}d).
The calculated forces are provided by a rigid CO molecule \textemdash that represents the oxygen atom at the apex\textemdash\space interacting with an stoichiometric CeO$_2$ slab (Fig.~S9), which are then converted to $\Delta f$~\cite{GiessiblPRB1997}. The long-range vdW interaction is obtained from a fitting to the experimental $\Delta f$ curves at large distances (cyan dotted line in Fig.~\ref{Fig1}d, 
see also Fig.~S9
and Methods).
The calculated
curves reproduce quantitatively the experimental spectroscopy, and allow us to 
estimate
an absolute value for the imaging distance (Z$_{\rm exp}=0$ in Fig.~\ref{Fig1}d),
which corresponds to a Z$\sim$411pm separation with respect to the plane defined by the centre of the uppermost surface oxygen atoms (Fig.~\ref{Fig1}d).
According to this distance and the $\Delta f$ curves, 
the bright contrast in the constant-height AFM images relates to the surface top-most oxygen atoms, and the spots with the lower contrast correspond to the position of the cerium atoms beneath them (see the inset models in Fig.~\ref{Fig1}d).

A force curve on top of the water molecule 
informs us that the molecule is imaged in Fig.~\ref{Fig1} past the
minimum of the $\Delta f$ curve; a region where the interaction between the probe and the surface has already become repulsive (see Fig.~S2 for a detailed evolution of the AFM contrast over the water molecule with the distance).
For surface atoms that correspond to the same chemical species, and with a similar charge state, the relative position of the $\Delta f$ minima could be an indication of 
their topographic distribution 
normal to the surface. The minimum at a water molecule is 200 to 300~pm above the surface oxygen atoms,
which is consistent with the results of our DFT calculations, where the oxygen atom of the molecule is 250~(220)~pm higher than the binding cerium atom for the molecular (hydroxyl pair) adsorption state.

Comparing with all previous scanning probe microscopy studies of water on ceria 
~\cite{Gritschneder_Water_2007, Torbruegge-ManipulationOfWaterMolecules, Thornton2010_Water}, %
the images in Fig.~\ref{Fig1} display two striking differences: (i),
the triangular feature extending over the three neighboring oxygen atoms surrounding the cerium site is now replaced by three well-defined lines pointing towards the oxygen atoms; and (ii),
these lines are no longer identical, with one of the prongs 
presenting a dimmer contrast than the other two, breaking the surface three-fold symmetry and giving rise to a boomerang-like shape (Fig.~\ref{Fig1}). 
This asymmetric shape is found in the three possible orientations compatible with the
surface symmetry 
(see Fig.~\ref{Fig1} and Fig.~S2). In all the cases, there is also a dark shadow 
enclosed by two of the arms and extending towards one of the coordination vacancy sites, but its location does not seem to correlate with the direction of the less intense arm.

While the 
shape of the water molecules 
reported here 
could be attributed to the higher resolution provided by the use of 
constant height imaging with a functionalized probe~\cite{Gross2009ChemicalStructureOfaMolecule}
and the very low temperature of the 
experiments,
the asymmetry
detected in our AFM images
poses a challenging question.
The low energy barriers among the different adsorption 
configurations~\cite{DeliaWaterOnCeria111}
were proposed to explain the
extended triangular contrast
reported in the previous experiments, leaving open the question of the true water adsorption state on CeO$_2$(111)
from an experimental perspective.
The clear asymmetry found now in our experiments suggests either the stabilization of the dissociated state in the form of a hydroxyl pair, that naturally breaks the symmetry in one of the directions, or the presence of defects like oxygen vacancies and
the associated Ce$^{3+}$ ions nearby, that locally change the structure and energetics of water adsorption. %

Water dosing in the 
experiments reported here
was carried out at room temperature (see Methods), resulting in a rather low coverage. 
Increasing the dosage of water in our CeO$_{2}$(111) films by 10 and 100 times did not provide a significant increase in the concentration of water on the surface.
According to Mullins et al., water molecules do not adsorb on a fully oxidized CeO$_{2}$(111) surface at room temperature~\cite{Mullins_WaterDissocoationCeO2111CeO2100}, suggesting that
the adsorption of water 
on the CeO$_{2}$(111) surface at room temperature is significantly affected by the presence of defects. This would be consistent with our early works on the observation and manipulation of water molecules on CeO$_{2}$(111) single crystals with AFM~\cite{Gritschneder_Water_2007, Torbruegge-ManipulationOfWaterMolecules}.
Those single crystals presented a high density of defects~\cite{Note_defects},
whose concentration and appearance greatly varied from one terrace of the crystal to another~\cite{Torbruegge_2007}. Those experiments, also carried out by similar dosing at room temperature, presented higher water coverage. 
Furthermore, while water was easily manipulated to a neighboring cerium site on the highly-defective CeO$_{2}$(111) single crystals~\cite{Torbruegge-ManipulationOfWaterMolecules}, our efforts to manipulate one of the water molecules displayed in Fig.~\ref{Fig1}a using a highly-reproducible method we reported before~\cite{Sugimoto05-NatureMat-LateralManip, Torbruegge-ManipulationOfWaterMolecules} did not succeed (see Fig.~S3).

\begin{figure}[h!]
\centering
\includegraphics[width=0.8\textwidth]{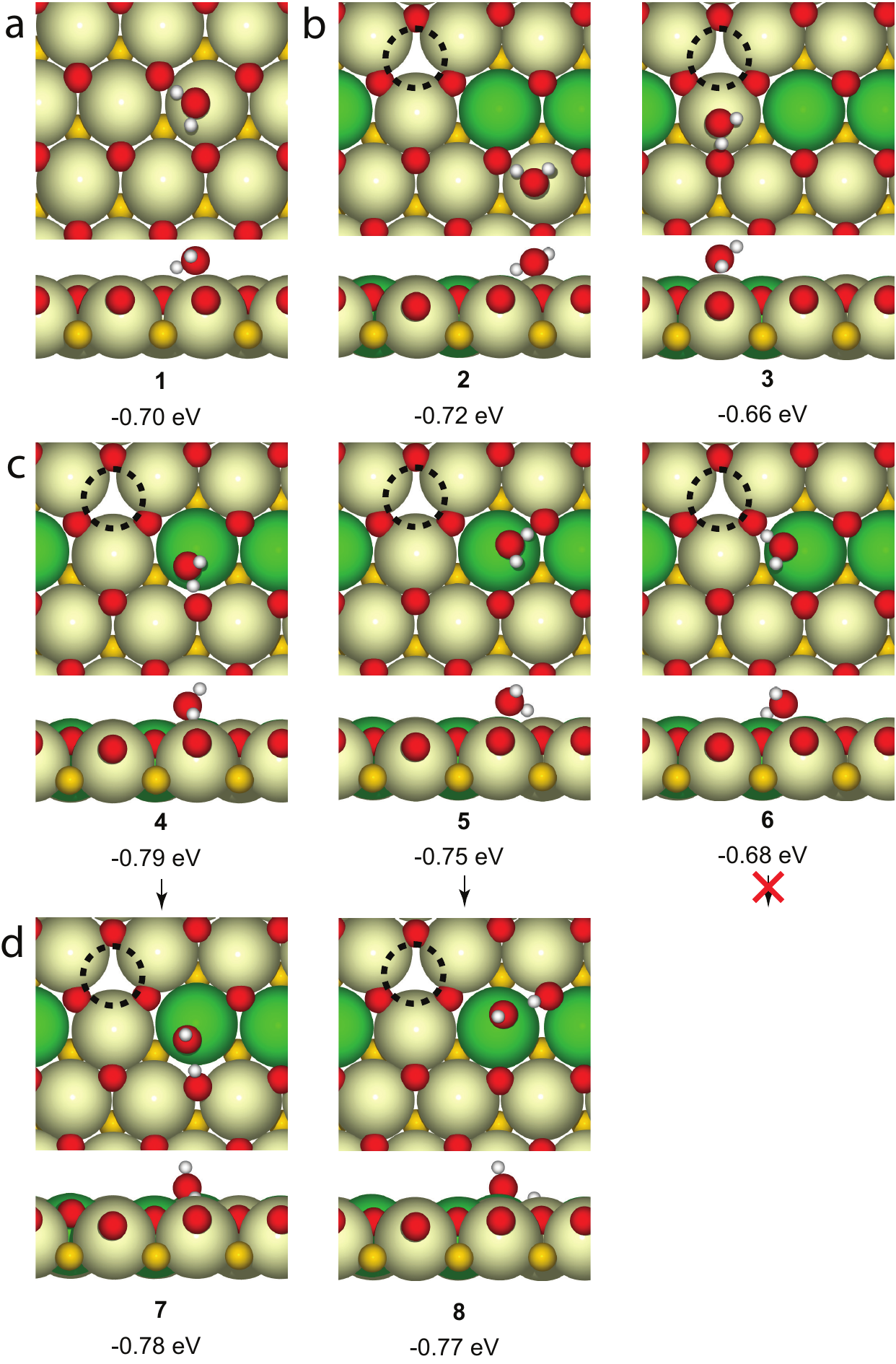}
\caption{{\bf Water adsorption at the CeO$_{2}$(111) surface in the presence of a subsurface oxygen vacancy.}
DFT optimized most stable structures for water adsorption on: {\bf a} a Ce$^{4+}$ of the pristine surface; {\bf b} a Ce$^{4+}$ of the reduced surface, far {\bf 2} and close {\bf 3} to the oxygen vacancy; and {\bf c} a Ce$^{3+}$ of the reduced surface. {\bf d} Hydroxyl pair (OH + H) adsorbed on a Ce$^{3+}$ and a neighboring O atom of the reduced surface. The energies with respect to the slab and water molecule energy are indicated. The position of the oxygen vacancy is highlighted by a dotted circle. Ce$^{3+}$, Ce$^{4+}$ and surface O and subsurface O atoms are represented in green, beige, red and yellow, respectively. No stable hydroxyl pair (OH + H) was obtained from structure {\bf 6}; it reverts to water.
}
\label{Fig2}
\end{figure}

To understand the possible influence of defects in the adsorption of molecular water on the CeO$_{2}$(111) surface, we have performed first-principles calculations based on DFT 
as implemented in VASP~\cite{Kresse1993Jan,Kresse1994May,Kresse1996Oct,Kresse1996Jul,Kresse1999Jan} using the PBE~\cite{Perdew1996Oct} exchange-correlation functional supplemented with long-range dispersion corrections described by the DFT-D3 approach~\cite{Grimme2010Apr,Grimme2011May} (see Methods).
We have previously demonstrated that the energetically most stable near-surface defect in the CeO$_{2}$(111) system is a subsurface oxygen vacancy (SSO$_V$), which generates lattice relaxation effects due to an excess-charge localization at two nearby Ce$^{3+}$ atoms~\cite{NiliusVeronica}.
Accordingly, in this work we have simulated the adsorption of a water molecule when a SSO$_V$ is located nearby using a 3$\times$3 unit cell.
Figure~\ref{Fig2}a reproduces our previous results on the adsorption of an individual water molecule on the pristine CeO$_{2}$(111) surface~\cite{DeliaWaterOnCeria111}, confirmed by later works~\cite{Rockert2020, Zou2023} %

In the presence of a nearby SSO$_V$, the adsorption energy of the water molecule on a Ce$^{4+}$ atom, -0.70 eV, is barely affected when the Ce$^{4+}$ is a second neighbor to the vacancy, -0.72 eV, but reduces (weaker binding) for first neighbors, with values in the range (-0.61, -0.66) eV (see Fig.~\ref{Fig2}b, Fig.~S5, and the associated discussion for a full account of all the calculated structures).
It is energetically more favorable for the water molecule to adsorb on a Ce$^{3+}$ than on a Ce$^{4+}$, -0.79 eV (Fig.~\ref{Fig2}c).
In this case, under the presence of a SSO$_V$, the threefold symmetry observed for the adsorption of water at the pristine surface~\cite{DeliaWaterOnCeria111} is broken, and an adsorption configuration with one of the hydrogen atoms interacting with a surface oxygen atom neighboring the SSO$_V$ becomes the less favorable of the three possible molecular orientations, -0.68 eV vs -0.75 and -0.79 eV (Fig.~\ref{Fig2}c).
A similar trend is observed when considering the hydroxyl pair: a lower adsorption energy is obtained over a Ce$^{3+}$ than over a Ce$^{4+}$, -0.78 eV vs -0.75 eV (see Fig.~\ref{Fig2}d, Fig.~S5 %
and associated discussion).
It should be noted that the hydroxyl pair cannot be formed with an oxygen atom neighboring the SSO$_V$; when forcing the hydrogen atom to bind such surface oxygen, the calculation recovers the molecular form (Fig.~\ref{Fig2}d).
The molecular and hydroxyl pair states for adsorption on a Ce$^{3+}$ are separated by energy barriers of $\sim 50-60$~meV (Fig.~S6b), smaller than those found for the adsorption on a Ce$^{4+}$ on the fully oxidized surface 80 (130)~meV for the transition from (to) the molecular state (Fig.~S6c). These calculations, 
using the DFT-D3 approach for the vdW interaction, compared well with our previous results~\cite{DeliaWaterOnCeria111} employing the optB86b-vdW functional, where binding energies of -0.73 (-0.76)~eV were found for the molecular (hydroxyl pair) states, with energy barriers of 80 (100)~meV
for the transition from (to) the molecular state.
The comparison with published theoretical results for water adsorption on a Ce$^{3+}$~\cite{Han_2021} shows that the inclusion of dispersion corrections provides an additional binding energy but does not modify significantly the relative stability of the different adsorption configurations for both the molecular state and the hydroxyl pair.

This symmetry breaking and the energetic diversity among otherwise similar adsorption configurations for the water molecule close to the vacancy are due to the structural surface relaxations induced by the presence of the SSO$_V$~\cite{NiliusVeronica} and the Ce$^{3+}$ ions.
Our calculations resulted in the oxygen atoms directly on top of the SSO$_V$ relaxing $\sim$15~pm towards the bulk. In contrast, the second nearest surface oxygen atoms and the nearby Ce$^{3+}$ rise $\sim$15~pm and $\sim$7~pm, respectively, toward the vacuum (Figs.~\ref{Fig6}d~and~e).
We have carefully checked that all of these conclusions regarding structure and energetics are not conditioned by the particular arrangement of the Ce$^{3+}$ ions forced by the 3$\times$3 cell (where they are neighbors), as shown by the results for adsorption structures (Fig.~S7) and energy barriers (Fig.~S8) obtained in a 4$\times$4 unit cell (see Methods).

\begin{figure}[b!]
\centering
\includegraphics[width=0.9\textwidth]{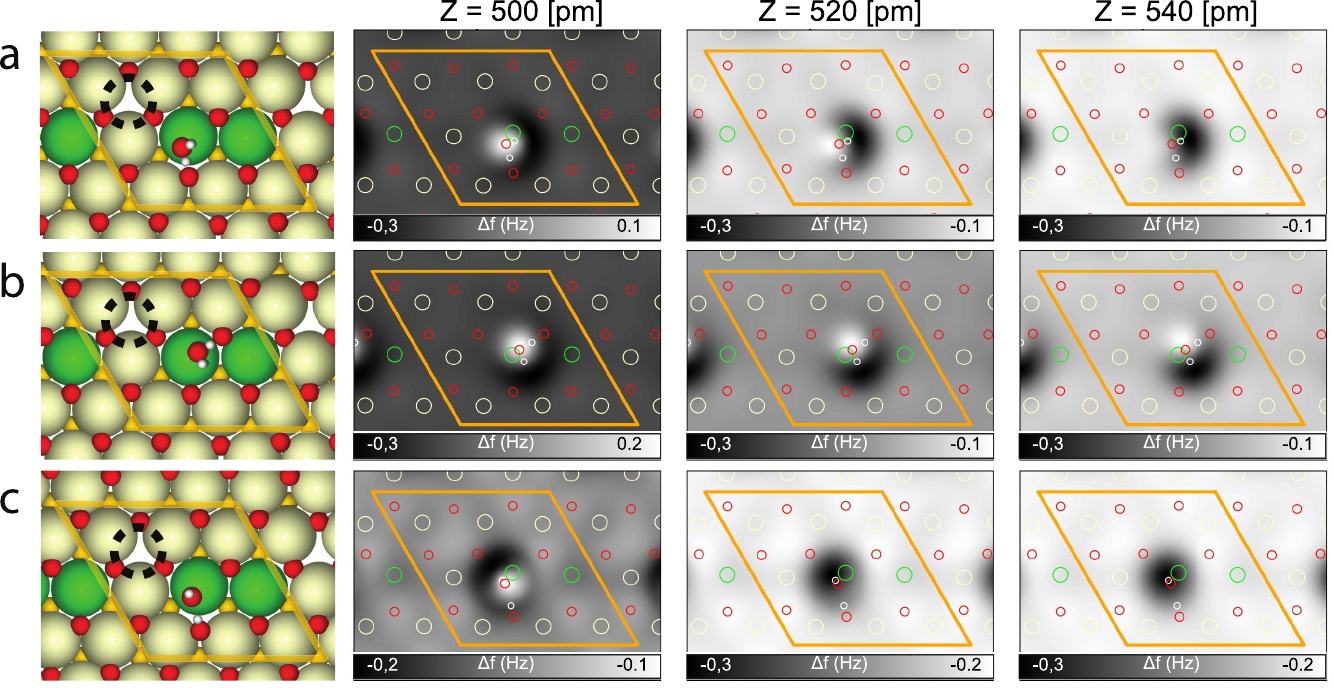}. %
\caption{{\bf Simulated 
constant-height AFM images.}
Calculated $\Delta f$ maps of the most stable adsorption configurations of a water molecule adsorbed on a partially reduced CeO$_{2-x}$(111) surface characterized by the presence of a subsurface oxygen vacancy (dotted circumference). The $\Delta f$ images were calculated using the Full Density Based Model (including the vdW interaction extracted from the experiments, see Methods) at three representative probe–surface separations for: {\bf a}, molecular water in configuration {\bf 4}; {\bf b}, molecular water in configuration {\bf 5}; and {\bf c}, hydroxyl pair in configuration {\bf 7}, as shown in Fig.~\ref{Fig2}. Latices with the top-most Ce and O surface atomic positions, as well as a rhomboid highlighting the unit cell, have been superimposed to each $\Delta f$ image. The color code for the species in the atomic models is the same as for Fig.~\ref{Fig2}.}

\label{Fig3}
\end{figure}

To explore whether the presence of the SSO$_V$ and associated Ce$^{3+}$ ions explain the asymmetric shape found in our experiments, we calculated constant-height AFM images at several probe-surface separations using representative stable configurations for the adsorption of water close to a SSO$_V$ (Fig.~\ref{Fig3}). 
These calculations are based on the Full Density--Based Model (FDBM)~\cite{Ellner2019, EVMAppSurfSci2023} \textemdash which retains DFT accuracy in the description of the 
probe-surface
forces while offering the computational speed needed for 
simulating a
whole image\textemdash\space using 
the probe model 
shown in Fig.~\ref{Fig1}. Therefore, the calculated $\Delta f$ signal for these images includes both the contribution from the CO molecule (calculated with FDBM) and the long-range vdW interaction extracted from the experiments.

Figures~\ref{Fig3}a~and~b present calculated $\Delta f$ images of the two most stable configurations for the water adsorption on a Ce$^{3+}$ atom in the molecular form ({\bf 4} and {\bf 5} in Fig.~\ref{Fig2}, respectively).
These orientations \textemdash which hold a hydrogen bond with a surface oxygen atom far from the SSO$_V$\textemdash\space show a bright contrast over a region close to the oxygen of the molecule, and display an extended dark feature over a wide area spanning from the position of the hydrogen atom pointing upwards 
towards the surface oxygen atoms opposite to the SSO$_V$.

The probe-surface separations explored (540~to~500~pm) covers the regime where the bright contrast over the oxygen atom of the molecule starts to be dominated by the Pauli repulsive interaction between the lone pairs of the oxygen atoms of CO and water; consistent with the appearance of a bright contrast \textemdash of repulsive nature, as it develops past the $\Delta f$ minimum\textemdash\space at the water site in Fig.~S2.
Analyzing the force contributions (Fig.~S10)
for the representative case of Fig.~\ref{Fig3}a reveals that the brightest area for the total force image (``static'' in Fig.~S10) corresponds to the highest contrast in the image displaying the Pauli repulsion (``SR'' in Fig.~S10) only for the closest separation (500~pm). For larger separations, the electrostatic interaction (``ES'' in Fig.~S10) also contributes to a degree in the calculated images.
The dark contrast reflects the attractive interaction associated with the onset of a hydrogen bond formation between the hydrogen pointing upwards and the oxygen atom at the CO molecule.

For the molecular adsorption configuration {\bf 5}, the calculated $\Delta f$ images (Fig.~\ref{Fig3}b), while presenting the same basic features discussed for configuration {\bf 4} (Figs.~\ref{Fig3}a~and~S10), there are subtle contrast differences ultimately motivated by structural surface relaxations introduced by the presence of the vacancy and the associated Ce$^{3+}$ ions.

For the adsorption of the water molecule as a hydroxyl pair (configuration {\bf 7} in Fig.~\ref{Fig2}d), the imaging mechanism 
is similar (Fig.~\ref{Fig3}c): an initial dark contrast is obtained due to an attractive interaction of the hydrogen atom pointing towards the vacuum with the oxygen atom at the probe, that develops into a bright spot upon further approach towards the surface, as the repulsive interaction between the oxygen of the hydroxyl and the one at the probe starts dominating the contrast (see Fig.~S11 %
for the corresponding force contributions).

These calculated $\Delta f$ images point towards the AFM being capable of discriminating between the two adsorption forms (molecular or hydroxyl pair) upon stabilization of one of them on the surface. However, the calculated images only partially reproduce the observed features in the experimental ones.
The calculated AFM images were obtained under the common assumption that the surface configuration is not modified by the interaction with the probe. 
Yet our DFT calculations suggest that water is extremely mobile in either of the two adsorption forms on CeO$_2$(111)~\cite{DeliaWaterOnCeria111}.
Further information about how the water molecule interacts with the probe can be obtained from the calculation of $\Delta f$ curves over the relevant atomic positions visited by the molecule, where the response to the presence of the probe is taken into account.
 
\begin{figure}[t!]
\centering
\includegraphics[width=0.45\textwidth]{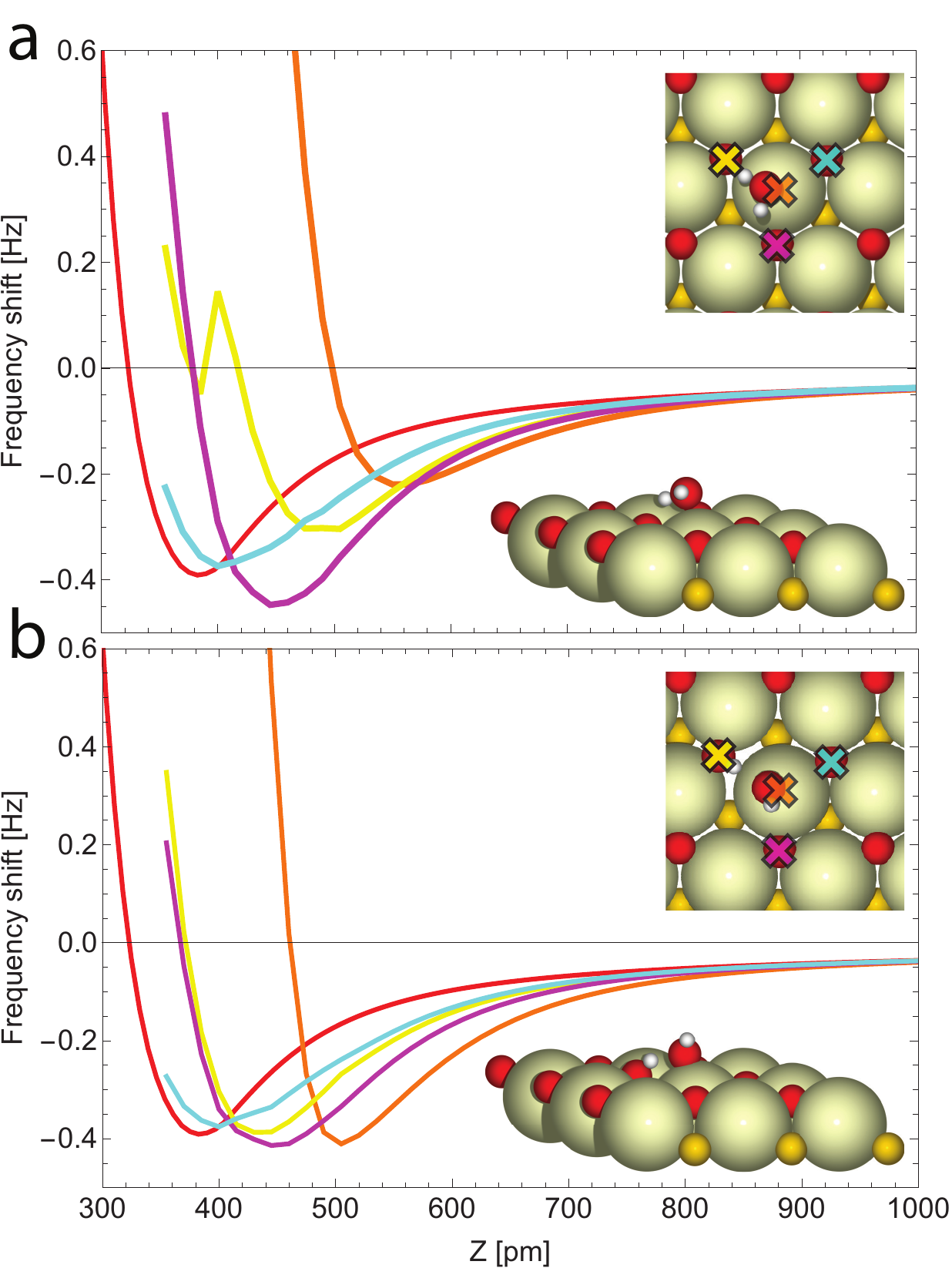}
\caption{{\bf Calculated force spectroscopy curves for a water molecule on a fully oxidized CeO{$_2$}(111) surface.}
Sets of calculated $\Delta f$ curves over the surface atoms visited by a water molecule adsorbed on the CeO{$_2$}(111) surface: the Ce atom the molecule binds to (in orange) and its three first-neighboring oxygen atoms (in yellow, cyan and magenta).
{\bf a}, Water in a molecular form; and {\bf b}, water in hydroxyl pair form.
The inset images are top and side views of the atomic conformations of the molecule and surface atoms.
The $\Delta f$ curve over a surface O atom in the absence of the water molecule (curve in red, Fig.~\ref{Fig1}d) has been included for reference. 
The calculation of the $\Delta f$ curve over the centre of each surface atoms (colored crosses) was performed to match the experimental counterparts (see Fig.~\ref{Fig5}). 
The probe model introduced in Fig.~\ref{Fig1}, including the long-range vdW interaction, was used for the calculation of the $\Delta f$ curves (see Methods).
}
\label{Fig4}
\end{figure}

Figures~\ref{Fig4}~a~and~b present calculated $\Delta f$ curves over the atomic sites of a fully oxidized CeO$_2$(111) surface for water adsorbed in molecular and hydroxyl-pair forms, respectively.
In the calculation of the forces that generate these curves (displayed in Fig.~S12), the outermost oxygen-cerium-oxygen trilayer of the surface and the water species were allowed to relax in response to the force exerted by the probe model introduced in Fig.~\ref{Fig1}.
Each set of $\Delta f$ curves comprises the interaction over the cerium site the water binds to (orange), the three neighboring surface oxygen atoms (yellow, cyan and magenta), and a surface oxygen atom of the fully oxidized surface (red) taken as reference.

The calculated $\Delta f$ curves for the adsorption of water in molecular form (Fig.~\ref{Fig4}~a)
reveal a strong variability in the magnitude and position of the minima over the neighboring surface oxygen atoms.
The water molecule pulls the oxygen atom involved in the hydrogen bond up by $\sim$10~pm with respect to the other two oxygen atoms, whose heights are unaffected by the presence of the molecule.
This structural modification is amplified by the water-probe interaction, causing the corresponding $\Delta f$ minimum to be shifted by $\sim 100$~pm towards the vacuum (yellow curve) with respect to the reference surface oxygen case.
The proximity of the water molecule also affects the position and magnitude of the $\Delta f$ minimum for the oxygen atom labeled in magenta, while the $\Delta f$ curve over the oxygen position far from the hydrogen atom (cyan) is the closest to the one of the surface oxygen taken as a reference.

The dominant role of the interaction versus structural effects can be clearly seen when considering the curve on the cerium site: although the oxygen in the water molecule is $\sim 170$~pm higher than the oxygen atom in the yellow site, the attractive interaction provided by the hydrogen atom pointing out of the surface reduces the difference in the position of the curve minimum to $\sim 75$~pm.
Despite the differences between the two systems (the hydrogen pointing out of the surface has a lower height than in the adsorption on a Ce$^{3+}$), the contrast evolution for the cerium site at large distances is quite consistent with the one shown by the images calculated with FDBM for the reduced surface, with the orange curve bending upward and crossing those associated with the neighboring oxygen atoms for distances below 540~pm.

The hydroxyl pair provides similar results, although differences among the three oxygen sites are significantly smaller. It is remarkable the effect of the greater flexibility of the atoms involved in the hydroxyl pair to move in response to the interaction with the probe, that changes significantly the position and strength of the $\Delta f$ minima for the cerium site and the oxygen site in yellow with respect to the molecular adsorption case.

\begin{figure}[h!]
\centering
\includegraphics[width=0.45\textwidth]{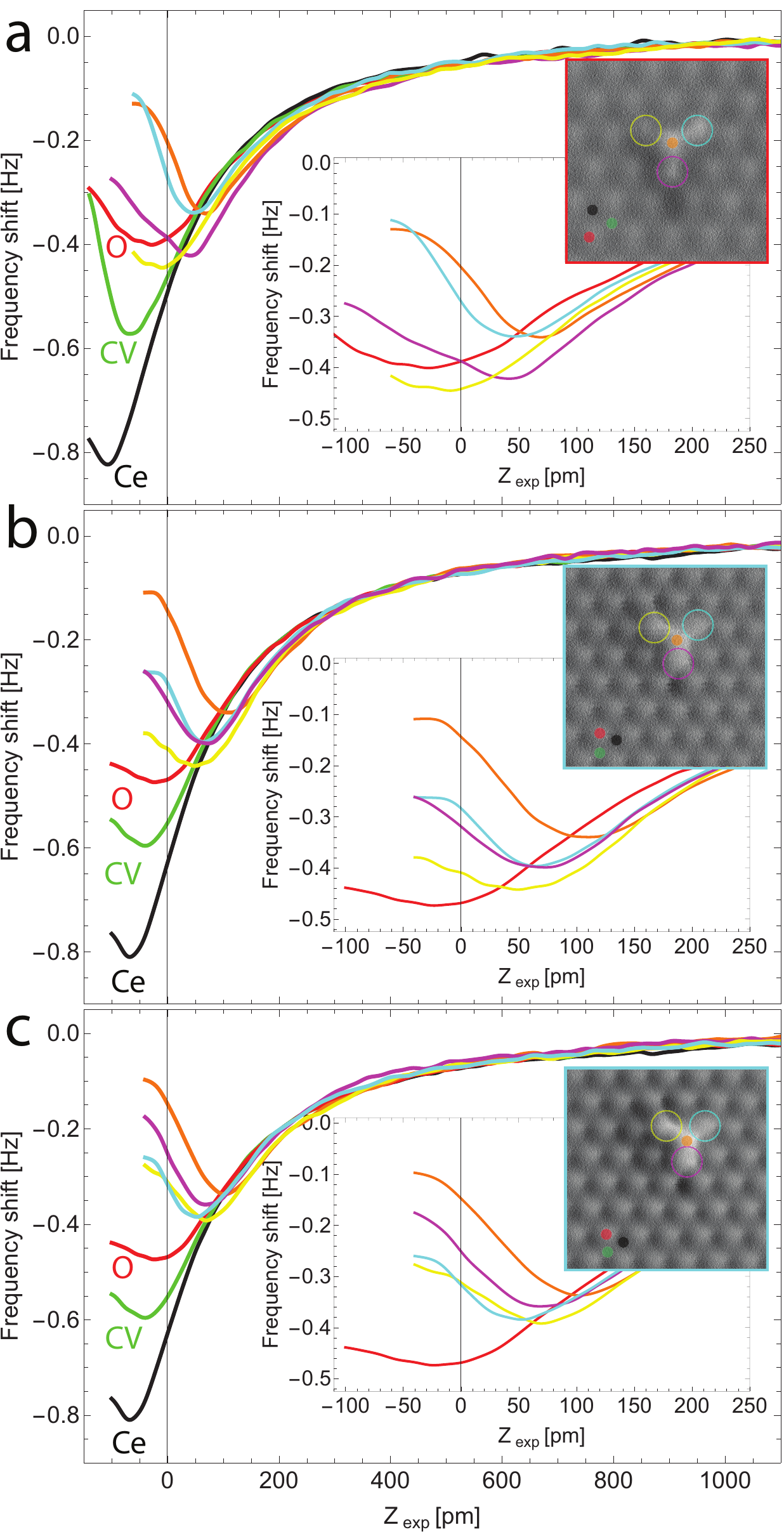}
\caption{{\bf Force spectroscopic measurements over water molecules.}
Frequency shift ($\Delta f$) curves obtained over the centre of the water molecule structure (orange) and the centre of the three surface oxygen atoms (yellow, cyan and magenta) the molecule visits. 
The inset graphs show a detail around the minima of the $\Delta f$ curves.
These sets of force curves were measured over:
{\bf a}, the molecule highlighted with a red square in Fig.~\ref{Fig1}a;
{\bf b}, the molecule highlighted with a cyan square in Fig.~\ref{Fig1}a;
{\bf c}, the same molecule as in {\bf b} but 40 minutes after the acquisition of the first set.
Constant height AFM images obtained prior the force spectroscopy measurements are included as insets.
Each image was measured at a probe-surface separation corresponding to the origin of the distance axis.
A set of force curves over the three surface sites of the CeO{$_2$}(111) surface (Ce in black, O in red, and CV in green) far from the water molecule are included for reference. 
Acquisition parameters were: $A$ =~60~pm and $f_{o}$=~994230~Hz.
}
\label{Fig5}
\end{figure}

Our experiments also reveal an asymmetric behavior of the $\Delta f$ curves obtained over the surface atomic sites visited by the water molecule. Figure~\ref{Fig5} presents three sets of curves measured over two different molecules:
Fig.~\ref{Fig5}a displays the curves obtained at the location highlighted by a red square in Fig.~\ref{Fig1}a; and Figs.~\ref{Fig5}b~and~c show two sets of curves measured over the molecule enclosed by a cyan square in Fig.~\ref{Fig1}a, but with a 40 minutes interval between their acquisition. The three sets are accompanied by $\Delta f$ curves measured over cerium, oxygen, and coordination vacancy surface sites far from the location of the water molecules. The inset graphs display details of the $\Delta f$ curves around the minima, including the reference surface oxygen one (in red).
All curves were measured over the centre of the corresponding atomic site and with identical probe apex termination.

A close look at the contrast of the atomic sites in the image displayed in Fig.~\ref{Fig5}a reveals a discrepancy with the spectroscopic data. Two of the surface oxygen atoms (highlighted in cyan and magenta) show a bright contrast, while the third one (yellow) has the appearance of a normal surface oxygen atom. The associated $\Delta f$ curves indicate, however, that the contrast over the spots in yellow and magenta should be almost identical to the contrast on a standard surface oxygen. The sets displayed in Figs.~\ref{Fig5}b~and~c show similar discrepancies: in Fig.~\ref{Fig5}b, the spot in yellow should have a dimmer contrast than the one in cyan; and in Fig.~\ref{Fig5}c, the sites highlighted in yellow and cyan should have almost identical contrast, but in the image the cyan spot appears dimmer than the yellow one.
This discrepancy between the image contrast and the corresponding $\Delta f$ curves, together with the variability of the $\Delta f$ curves recorded at different times over the same molecule (Figs.~\ref{Fig5}b~and~c) point to the existence of probe-induced dynamical effects that result in the change of the adsorption configuration of the water molecule
around the cerium atom, and possibly, reversible transitions between molecular and hydroxyl pair forms.

These probe-induced changes have been verified during the simulations to produce the calculated $\Delta f$ curves in Fig.~\ref{Fig4}.
A striking feature appearing in these curves is an abrupt instability upon approaching the probe beyond 420~pm over the oxygen atom at the yellow site, that brings the $\Delta f$ curve very close to the one at the magenta site (Fig.~\ref{Fig4}a). This instability originates from a probe-induced change in the adsorption configuration, in which the hydrogen atom interacting with the oxygen at the yellow site moves upward, breaks the hydrogen bond, and then the whole molecule rotates to establish a new hydrogen bond with the oxygen atom at the magenta site (see Movie~S1). In this new arrangement, the yellow site plays an equivalent role to the magenta site in the original configuration, explaining the similarity between the two $\Delta f$ curves beyond 400~pm approach. 
This change in the adsorption configuration also takes place when the 
probe
approaches other sites, like the 
coordination vacancy
between the
oxygen atoms labeled in yellow and cyan
(see Movie~S2), confirming that this process 
\textemdash characterized by a very low energy barrier~\cite{DeliaWaterOnCeria111}\textemdash\space
is quite common for the probe-surface separations 
used in 
atomic resolution
imaging ($\sim 411$~pm for the images in Fig.~\ref{Fig5}).
Interestingly, the
probe is also able to 
induce the transition from molecular form
to hydroxyl pair when approaching over the cerium site (see Movie~S3).
In this case, the proton transfer occurs spontaneously \textemdash indicating a zero energy barrier\textemdash\space in the range of 385--400~pm, close to the imaging separation mentioned above.
This transition will seamlessly switch between the orange curves in Figs.~\ref{Fig4}a~and~b, bringing the position of the minimum \textemdash relative to the ones for the oxygen sites\textemdash\space and its strength closer to the experiment.
The separation at which this transition occurs is well passed the minimum in the calculated $\Delta f$ curves, entering a probe-surface interaction range dominated by the repulsion between the oxygen atoms of molecule and probe.
The limitation of our model to describe areas of highly repulsive forces, where the long-range elasticity of 
probe and surface
starts playing a role, 
makes the theoretical curves significantly steeper than the experimental ones
past the minimum of the $\Delta f$.
However, considering that the water molecule is imaged in the experiments 
under this repulsive interaction regime (orange curves in Fig.~\ref{Fig5} and Fig.~S2), our calculations strongly suggest that the transition between molecular form and hydroxyl pairs should take place systematically during imaging.

With all this information about the atomistic behavior of the water molecule on a partially reduced CeO$_{2-x}$(111) surface, and its interaction with the probe during AFM imaging and spectroscopy, we can address the appearance of the molecule as three well-defined lines pointing to the neighboring oxygen atoms with a boomerang-like shape.

When imaging with CO-functionalized probes, it is well established that the relaxation of the CO molecule in response to the interaction with the surface enhances the AFM signal over covalent~\cite{Gross2009ChemicalStructureOfaMolecule, HapalaPRB2014} as well as hydrogen~\cite{HapalaPRB2014, Ellner2019} bonds, which are imaged as bond-like sharp features.
In our experiments and simulations we use a rigid oxygen-terminated apex, yet the probe-induced transitions of the water molecule between the different adsorption configurations play a similar role to the relaxation of the CO molecule mentioned above, leading to the three well-defined lines observed in our experimental images. These transitions are not privative of the molecular form, but they also happen for the hydroxyl pair. Movie~S4 shows a probe-induced change in the orientation of the hydroxyl with respect to the hydrogen bond, 
which closely resembles the dynamics of a flexible CO-apex 
in the presence of a 
rapidly varying region of the potential energy surface \textemdash a bond, for instance\textemdash\space that produces the sharp features in the AFM images.

These probe-induced relaxations would tend to restore the three-fold symmetry, leaving the origin of the asymmetry that generates
the boomerang-like shape %
an open question.
However, the discussion above shows that both the presence of defects \textemdash such as a SSO$_V$\textemdash\space
and the water molecule itself \textemdash that pulls up a surface oxygen atom with the hydrogen bond\textemdash\space contribute to break the symmetry among the three neighboring oxygen sites. 
A quantitative
answer
to this question would imply DFT simulations of force
curves
for a water molecule adsorbed on different cerium sites on a reduced ceria surface on a fine 3D grid of the 
probe
positions. 
The inclusion of just a single SSO$_V$ with the two associated Ce$^{3+}$ ions in those calculations renders the relaxation of the whole system to the ground state for each probe position a herculean feat, as the Ce$^{3+}$ ions make the electronic convergence quite cumbersome.
Such
calculation is beyond our current capabilities, but we can point out some of the factors that can contribute to the asymmetry. The most relevant is the differences among the oxygen sites induced by the presence of a SSO$_V$,
introduced 
in Fig.~\ref{Fig2}. 
This situation singles out the oxygen site closer to the vacancy, making less probable
the formation of 
a hydrogen bond or an hydroxyl with that site
of the countless times the water molecule visits the three oxygen atoms during an AFM image.

In the experiments reported here, direct evidence of the presence of subsurface oxygen vacancies could not be obtained either in the $\Delta f$ or in the dissipation signal; the latter is probably due to the use of a much stiffer force sensor~\cite{Sugimoto_KolibriSpringConstant} than in previous works~\cite{Torbruegge_2007}. However, we often observe cerium atoms with an unusual contrast (see the yellow arrows in Fig.~\ref{Fig1}a) in a concentration close to the coverage of water molecules on the terraces after dosing water at room temperature. 
We propose that these unusual, 
more attractive
cerium atoms can be candidates for Ce$^{3+}$, originating from the presence of defects under the surface.

\begin{figure}[t!]
\centering
\includegraphics[width=0.9\textwidth]{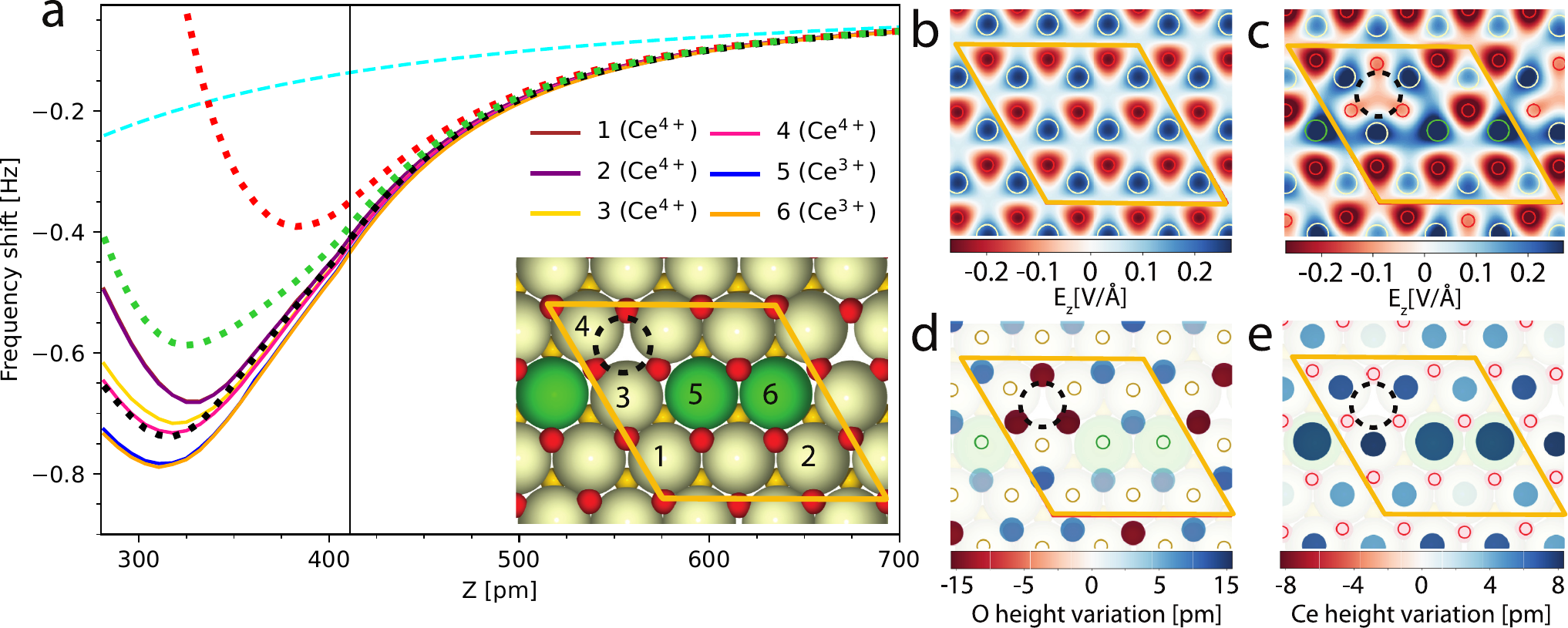}
\caption{{\bf Calculated frequency shift 
curves 
over cerium atoms in the presence of a vacancy.}
 {\bf a}, DFT-based calculated $\Delta f$ over the Ce$^{4+}$ (curves 1-4) and the Ce$^{3+}$ (curves 5 and 6) atoms of a reduced CeO$_{2-x}$(111) surface in the vicinity of a subsurface oxygen vacancy (dotted-line circles).
Curve labeling is indicated in the 
inset image displaying the atomic arrangement, which also serves as reference for the atomic features presented in the other panels.
The $\Delta f$ curves for the O (red), Ce (black), and CV (green) calculated for the fully oxidized surface, together with the long-range vdW component (cyan) already displayed in Fig.~\ref{Fig1}, have been included as dotted lines for reference.
This long-range vdW contribution, extracted from the experiments, is incorporated in all the $\Delta f$ curves in the graph.
{\bf b} and {\bf c}, Local distribution of the vertical component of the electric field ($E_{z}$) for the fully oxidized surface and for the same area with a subsurface oxygen vacancy (dotted circle), respectively, both obtained at a height of 274~pm above the topmost oxygen layer.
{\bf c} Vertical relaxation of the surface oxygen atoms near a subsurface oxygen vacancy with respect to the position of the topmost oxygen layer of the fully oxidized surface.
{\bf d} Vertical relaxation of the cerium atoms in the proximity of a subsurface oxygen vacancy with respect to the position of the outermost cerium layer of the fully oxidized surface (78~pm below the uppermost oxygen layer).
The parallelogram in 
orange
highlights the unit cell used in the calculations.
}
\label{Fig6}
\end{figure}

To address the possible nature of these unusual cerium atoms, 
we calculated 
$\Delta f$ curves on different cerium sites of the reduced surface (Fig.~\ref{Fig6}a) using the model probe introduced in Fig.~\ref{Fig1}, in a similar fashion as for Fig~\ref{Fig4} (see Methods).
However, in order to speed up 
the
DFT-base calculation of the forces (Fig.~S13),  
and more importantly, to single out the effect of the vacancy presence, we have not included the water molecule in these simulations.

Our theoretical results reveal clear differences between the Ce$^{3+}$ and Ce$^{4+}$ sites, as well as among the Ce$^{4+}$ atoms themselves, depending on their proximity to the SSO$_V$. The Ce$^{4+}$ atoms adjacent to the SSO$_V$ present a deeper minimum 
than the Ce$^{4+}$ atoms far from it. The Ce$^{3+}$ sites have the deepest  
$\Delta f$
minimum
(18\% larger than the Ce$^{4+}$ labeled 1, and 9\% larger than the reference Ce$^{4+}$ on the fully oxidized surface), which is associated with a significantly stronger attractive force with the probe compared to the Ce$^{4+}$ sites (see Fig.~S13). %

This outcome seems counter-intuitive when considering the extra screening provided by the additional electron localized at the Ce$^{3+}$ site.
However, an increase in the local electric field over the Ce$^{3+}$ as a result of surface atomic relaxations accounts for this stronger interaction.
Figures~\ref{Fig6}b~and~c show the vertical component of the local electric field ($E_z$) for the fully oxidized and the reduced ceria surface, respectively, at a height of 274~pm above the uppermost oxygen layer. The Ce$^{3+}$ sites clearly have the strongest $E_z$, followed by the Ce$^{4+}$ sites close to the SSO$_V$. 
This enhancement of the $E_z$ comes from distortions in the oxygen lattice associated with the excess charge localized in the Ce$^{3+}$ as a polaron.
These distortions displace the nearby oxygen atoms away from the Ce$^{3+}$ site (see Fig.~S14), thereby 
enhancing the attractive short-range electrostatic interaction and reducing the Pauli repulsion
sensed by the AFM probe upon approach.  %
The downward displacement of the oxygen atoms around the SSO$_V$ (Fig.~\ref{Fig6}d) and the slight upward shift of the Ce$^{4+}$ sites close to it (Fig.~\ref{Fig6}e) contribute to the enhanced $E_z$ and the stronger attraction compared to the Ce$^{4+}$ sites far from the SSO$_V$.
These simulations set the basis for the possibility of identifying the presence and location of the elusive Ce$^{3+}$ ions using spectroscopy measurements, 
supporting our interpretation for the cerium atoms with an unusual contrast found in the terraces.

\begin{figure}[t!]
\centering
\includegraphics[width=0.45\textwidth]{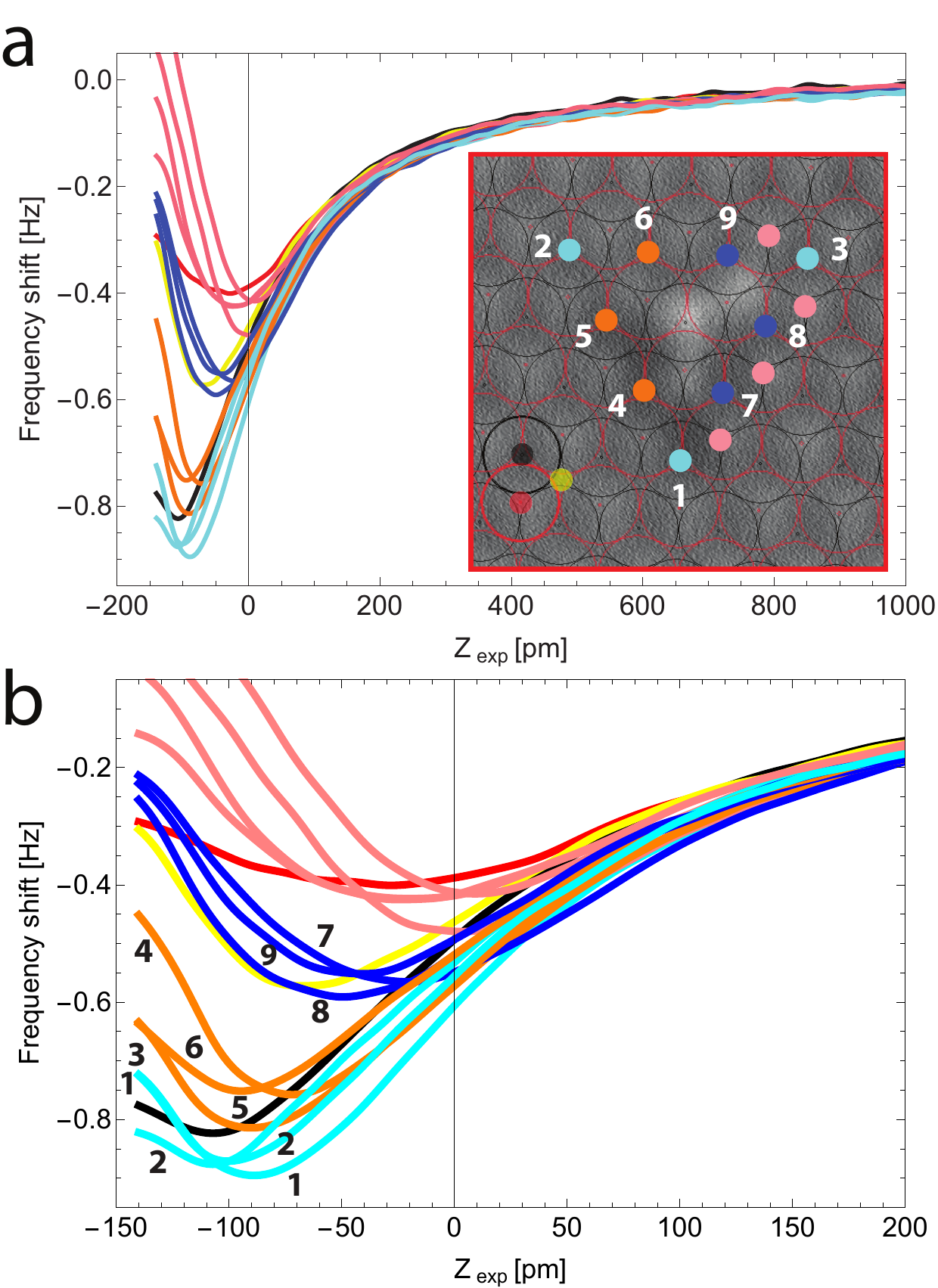}
\caption{{\bf Force spectroscopy measured at surface atomic sites surrounding a water molecule.}
{\bf a}, Sets of $\Delta f$ curves measured over Ce and O positions around the water molecule highlighted with a red square in Fig.~\ref{Fig1}a.
The inset is a constant-height AFM image of the molecule, acquired before the spectroscopic measurement at a separation corresponding with the origin of the distance axis.
Black and red lattices superimposed to the image highlight the cerium and oxygen surface positions, respectively.
{\bf b}, Detail of the $\Delta f$ curves close to the minima.
Numbering and color link the corresponding curve to the acquisition location in inset image: curves over the Ce atoms are in cyan (1, 2, 3), orange (4, 5, 6) and blue (7, 8 and 9); and relevant curves over O atoms are displayed in pink color (see also Fig.~S11).
Curves measured over Ce (in black), O (in red) and CV (in yellow) sites far from the water molecule are included for comparison.
Acquisition parameters were: $A$ =~60~pm and $f_{o}$=~994230~Hz.
}
\label{Fig7}
\end{figure}

In the experiments, we observed similar variability in the interaction with the cerium atoms in the vicinity of the water molecules.
Figure~\ref{Fig7} summarizes $\Delta f$ curves measured over atomic positions around the molecule highlighted with a red square in Fig.~\ref{Fig1}a. Additional details can be found in Fig.~S4. %
In view of the interaction with the probe, the curves over the cerium sites can be classified into three groups.
The cerium atoms at the vertices of the triangle formed by the water structure (curves 1, 2, 3 in cyan) show $\sim$9\% and $\sim$6\% greater minimum value than the one of the reference cerium atom far from the molecule (black curve).
The $\Delta f$ curves measured over the six cerium sites that first neighbor the water-occupied cerium, which should be equivalent by symmetry, show two distinctive behaviors. 
The three cerium sites on the left of the image (curves in orange) show similar response as the reference cerium atom, with the curve in position 4 being slightly more steep pass the minium, which is a sign of a stronger repulsive interaction taking place. For the three cerium sites on the right (curves 7, 8, 9 in blue), the interaction is clearly far more repulsive past the minium than the rest of the cerium atoms.
Similar strong repulsion is also detected over the second line of surface oxygen atoms on the right (positions and curves in pink).
The water molecule visiting more often the oxygen atoms on the right and the lower sides than the left one undoubtedly plays a role in adding steepness to the $\Delta f$ curves beyond the minima over the cerium atoms in blue, the oxygen sites in pink and also, in less degree, the cerium at position 4.
However, the cerium atoms at positions 1, 2, 3 (in cyan) are far enough to be affected by the water molecule, and the consistency with the theoretical calculations, points them out as candidates to Ce$^{3+}$ ions, suggesting the possible presence of SSO$_V$ close to the adsorbed water molecule.

At this point, we cannot rule out the presence of a SSO$_V$ close to the cerium atoms with an unusual strong attractive interaction in both the middle of the terraces and around the water molecules, but we speculate, considering the difficulty in increasing the water coverage on the surface when dosing at room temperature, that the presence of one or several SSO$_V$ close to the Ce$^{3+}$ ions is needed to bind water molecules to the terraces of the CeO$_2$(111) surface at room temperature.
Addressing this and other questions will require future work, exploiting the sensitivity of AFM spectroscopy in combination with tunneling measurements~\cite{NiliusVeronica} that could help locating the SSO$_V$s.

In summary, we have combined Non-Contact Atomic Force Microscopy and first--principles calculations to investigate the interaction of water with a partially reduced CeO$_{2-x}$(111) surface at the atomic scale. 
Our low--temperature experiments using oxygen--terminated probes reveal water molecules adsorbed on cerium atoms as sharp, asymmetric boomerang--like features, in contrast with previous STM and AFM studies, in which water appeared as a triangular shape, consistent with the symmetry of the lattice surface.
Density Functional Theory (DFT) simulations reveal that water \textemdash in both molecular and dissociated forms\textemdash\space preferentially adsorbs at Ce$^{3+}$ sites near subsurface oxygen vacancies, and the surface relaxations induced by these defects break the local symmetry of the lattice.
Calculated force spectroscopy based on DFT exhibits distinct interaction signatures between the Ce$^{3+}$ and Ce$^{4+}$ sites. These differences are attributed to local structural relaxations of surface oxygen atoms, which change the local electrostatic environment of the cerium atoms. These calculations match quantitatively with experimental force spectroscopy measurements around the water molecule.
Finally, the characteristic contrast in constant-height AFM images originates from a probe-induced mobility of the water molecule, enabled by low-energy barriers between different adsorption configurations and adsorption states, and possibly, from the particular environment created around a Ce$^{3+}$ ion close to a subsurface oxygen vacancy.
Beyond resolving water adsorption on the CeO$_2$(111) surface, our work demonstrates the broader capabilities of chemically selective AFM for probing local chemical reactivity. This approach offers a unique opportunity to investigate, at the atomic level, complex catalytic systems including single-atom catalysts, metal–support interfaces and defect-engineered oxide surfaces, where local structure and charge state critically influence chemical behavior.

\section*{Methods}\label{Methods}
\subsection*{Scanning probe microscopy experiments}\label{Exp}

The experiments were carried out in an ultra-high vacuum (UHV) system equipped with tools for the in situ sample preparation and a home-made scanning probe microscope operated at 4.8~K using a commercial controller (Nanonis SPM Control System, SPECS, Germany).
AFM experiments were carried out using the frequency modulation detection scheme~\cite{Albrecht91-J.Appl.Phys.-FreqModulation} keeping constant the oscillation amplitude of the force sensor. Under this scheme, the AFM signal corresponds to the shift in the free-oscillation resonant frequency of the force sensor ($\Delta f$) upon forces acting on it.
Force spectroscopy experiments were performed by initially approaching the probe towards the surface from the imaging separation, and then retracting it upon reaching the specified closest approach until reaching the free-oscillation regime.
Both constant-height AFM imaging and spectroscopy were carried out by setting voltage bias to zero.
CeO$_{2}$(111) thin-films were grown on a Cu(111) surface following a similar procedure to the one described elsewhere~\cite{SUTARA20086120}.
The Cu(111) single crystal was cleaned by repeated cycles of argon ion sputtering and annealing in UHV, and then the surface was oxidized by keeping the sample at 470~$^{\circ}$C in a $1.0 \times 10^{-5}$~Pa oxygen environment for 10~min. 
The CeO$_{2}$(111) thin-films were fabricated by depositing cerium (rod, 99.9~\% purity, GoodFellow) from a water-cooled e-beam evaporator in the presence of $1.0 \times 10^{-5}$~Pa oxygen while keeping the pre-oxidized Cu(111) sample at 480~$^{\circ}$C.
After the cerium evaporation, the sample temperature was gradually reduced at a typical rate of 1~$^{\circ}$C/sec in an $1.0 \times 10^{-5}$~Pa oxygen.
Upon the sample reaching 200~${\circ}$C, both heating and oxygen flow were stopped and the crystal was left to cool to room temperature in UHV.
Ultra-clean Milli-Q water, further purified by several freeze-pump-thaw cycles, was dosed by backfilling the UHV preparation chamber with water vapour through a leak valve while keeping the sample at 35$~{\circ}$C.
Dosing with 5, 50 and 500 Langmuir of water with the sample at 35$~{\circ}$C produced similar coverages as the one displayed in Fig.~\ref{Fig1}.

\subsection*{Probe preparation}\label{probe_preparation}
We used the KolibriSensor (SPECS, Germany) for the detection of both tunneling current and probe-surface interaction forces.
The probe of the KolibriSensor was sharpened to a typical apex radius of $\sim$15~nm ex situ by using a focused ion beam, and it was further conditioned for atomic resolution imaging in situ on copper oxide surface areas coexisting with the CeO$_{2}$(111) thin-films.
Probe conditioning is carried out until good and sharp atomic resolution is obtained and force spectroscopy over the three sites of the CeO$_{2}$(111) surface reproduces the trend shown in Fig.~\ref{Fig1}d.
Measurements were done on the fourth ceria surface bilayer (see Fig.~S1), in which an atomic arrangement close to the bulk ceria is expected.

\subsection*{Density Functional Theory (DFT) Calculations}

The adsorption and transition state structure optimizations, as well as the simulated force-distance calculations, were performed using Density Functional Theory (DFT) as implemented in the VASP code (version 5.4.4)~\cite{Kresse1993Jan,Kresse1994May,Kresse1996Oct,Kresse1996Jul,Kresse1999Jan} with the slab-supercell approach~\cite{Payne1992Oct}. 
The projector augmented wave (PAW) method~\cite{Blochl1994Dec} was used to describe the valence electrons of the atomic species: Ce (4f, 5s, 5p, 5d, 6s), O (2s, 2p) and H (1s), with a plane-wave cutoff energy of 415~eV. 
The electron localization on Ce$^{3+}$ atoms of the support has been treated by means of the DFT+U approach proposed by Dudarev et al.~\cite{Dudarev1998Jan}, with a $U_{eff}$ value of $U - J = 4.5$~eV for the Ce 4f electrons.
We used the PBE exchange--correlation functional, the generalized gradient approximation (GGA) suggested by Perdew, Burke, and Ernzerhof (PBE)~\cite{Perdew1996Oct}. 
Long-range dispersion corrections were considered by means of the so-called DFT-D3 approach~\cite{Grimme2010Apr,Grimme2011May}.

\subsubsection*{Adsorption and transition state structure calculations}

CeO$_2$(111) surfaces with (3$\times$3) and (4$\times$4) periodicities were modeled with an optimized lattice constant of 5.485~\AA\ for bulk CeO$_2$. 
The (3$\times$3) surface models have four O--Ce--O tri-layers, whereas the (4$\times$4) models were built with only three tri-layers to save computational resources.
Both models have sufficient separation between consecutive slabs ($\sim$12~and~27~\AA, respectively). %
All atoms in the bottom O--Ce--O tri-layer were kept fixed at their optimized bulk-truncated positions during geometry optimization, whereas the rest of the atoms were allowed to fully relax.
A (2$\times$2$\times$1) k-point mesh, according to the Monkhorst--Pack method~\cite{Monkhorst1976Jun}, was used to sample the Brillouin zone for the (3$\times$3) models, and only the $\Gamma$-point for the (4$\times$4) surface. For the gas-phase calculations of the water molecule, a (15$\times$15$\times$15) \AA$^3$ cell was employed, with $\Gamma$-point only.

Two Ce$^{4+}$ are reduced to Ce$^{3+}$ when an oxygen vacancy is generated.
Following the previous work by Sauer et al.~\cite{Pan2013Nov} (which reported many different positions for the Ce$^{3+}$ on a (4$\times$4) surface at the HSE level~\cite{Heyd2003May, Krukau2006Dec}), we modelled the most stable configuration they found, where the Ce$^{3+}$ are next-nearest neighbors of the vacancy (Fig.~S5a). 
Note that this causes the two Ce$^{3+}$ atoms to be neighbors (to each other) on the (3$\times$3) model (Fig.~S5a).
To check that our results on the stability of water and OH+H species were not significantly affected by this situation, and to ensure that the (3$\times$3) results were reliable for the subsequent spectroscopic studies, we also calculated the crucial structures on the (4$\times$4) model for the two most stable relative positions of the two Ce$^{3+}$ atoms (Figs~S5b~and~S5c).

The adsorption energy of water was calculated as:
\begin{equation}
\Delta E_{ads} = E(\text{H}_2\text{O}/\text{CeO}_2(111)) - E(\text{H}_2\text{O}) - E(\text{CeO}_2(111))
\end{equation}
where $E(\text{H}_2\text{O}/\text{CeO}_2(111))$ is the energy of the structure with water adsorbed on the surface, $E(\text{H}_2\text{O})$ is the energy of the molecule in the gas phase, and $E(\text{CeO}_2(111))$ is the total energy of the clean model surface.

To identify transition state structures, the climbing image nudged elastic band technique (CI-NEB) was used~\cite{Henkelman2000Dec}, and frequency calculations were employed to check that only one imaginary frequency was found for each of them (Figs~S7b-c~and~S9a-c).
Activation energies ($\Delta E_{act}$) are defined as the difference between the energy of the transition state (TS) and the initial state (IS).

\subsubsection*{Simulated frequency shift curves}

The frequency shift curves were calculated from the forces obtained with a model probe that includes a CO molecule to describe the interaction of the sample with the oxygen-termination of the probe apex plus an attractive long-range term that represents the vdW interaction from the mesoscopic part of the probe.
The interaction with the apex as a function of the probe height is calculated with DFT by vertically approaching a rigid CO molecule (no relaxations allowed) toward the surface, adding up the vertical component of the force over the two atoms of the CO molecule at each approach distance. The atoms of the water molecule and the first trilayer of the surface were allowed to relax at each step of the approach, while the remaining atoms in the surface were kept fixed. These relaxations are minimal for the calculation of force curves over sites of the bare ceria surface within the probe-surface distance range explored, and can safely be disregarded to speed up calculations, but play a crucial role in the description of the water response to the probe-sample interaction.
The computed forces curves were converted into frequency shift curves using classical perturbation theory~\cite{GiessiblPRB1997} and the efficient approach proposed by Giessibl~\cite{Giessibl2001APL}.

To include the Van der Waals (vdW) interaction between probe and surface, we fitted the experimental curves
considering the analytical expression for the frequency shift ($\Delta f$) caused by the Van der Waals interaction force, $F = - \frac{C}{\left(z_{ts}\right)^{2}}$, between a spherical probe and a plane~\cite{Garcia2002Sep}:
\begin{equation} 
\Delta f_{VdW} = \frac{f_0}{\sqrt{8 \pi} \cdot k_0 \cdot A^{3 / 2}} \cdot \frac{C}{\left(z_{ts}\right)^{3 / 2}}, 
\end{equation}
where $f_0$, $k_0$, and $A$ represent the free-oscillation resonant frequency, the stiffness,
and the oscillation amplitude of the force sensor, respectively, $z_{ts}$ denotes the probe-surface distance, and $C$ is a fitting parameter set to $C = -1.4311 \times 10^{-6}~$N pm$^2$ to match the experiments. %
\\

\subsection*{Simulation of the constant-height AFM images}

Constant-height AFM images were simulated using the Full Density Based Model (FDBM)~\cite{Ellner2019, Zahl2021Nov, Ventura-Macias2023Oct} as implemented in the
\texttt{DBSPM} github repository: \href{https://github.com/SPMTH/DBSPM}{\texttt{https://github.com/SPMTH/DBSPM}}.
This method efficiently computes frequency shift images while preserving the accuracy of the DFT-calculated forces, enabling the high-fidelity reproduction of atomic resolution AFM images.
FDBM calculates the total force between a probe with an inert termination, like a CO molecule, as the sum of three contributions: the short-range Pauli repulsion (SR), the electrostatic interactions (ES), and the Van der Waals dispersion (VdW).
The SR and ES contributions were computed as:
\begin{equation} 
V_{SR} = V_0 \int [\rho_{\text{probe}} \cdot \rho_{\text{sample}}]^{\alpha} \, dV,
\end{equation}
\begin{equation} 
V_{ES} = \int \rho_{\text{probe}} \cdot \Phi_{\text{sample}} \, dV, 
\end{equation}
where $\rho_{\text{probe}}$ and $\rho_{\text{sample}}$ represent the charge density of probe and sample, respectively, and $\Phi_{\text{sample}}$ is the electrostatic potential of the sample.
These quantities were obtained from independent DFT calculations. 
The VdW interactions were modeled using the DFT-D3 method~\cite{Grimme2010Apr}.
The parameters $\alpha = 1.08$ and $V_0 = 36.96$ eV/\AA$^{3(2\alpha-1)}$ were optimized to match DFT-based force spectroscopic curves above the reduced CeO$_{2-x}$(111) surface within the experimental probe-sample distance range.
The long-range vdW term (that accouns for the interaction with the mesoscopic part of the probe) described above was also added to the total force.
The total forces computed over a fine grid were converted into constant-height frequency shift images using classical perturbation theory~\cite{GiessiblPRB1997} and the efficient approach proposed by Giessibl~\cite{Giessibl2001APL}.
The FDBM simulated images shown in Fig.~\ref{Fig3} were performed for water adsorbed on the reduced CeO$_{2-x}$(111) surface.

\bmhead{Supplementary information}

\begin{itemize}
\item File containing figures S1 to S14 and related description. 
\item Movie S1: Simulation of a probe-induced translation of a water molecule in molecular form approaching over surface oxygen atom.
\item Movie S2: Simulation of a probe-induced translation of a water molecule in molecular form approaching over coordination vacancy.
\item Movie S3: Simulation of a probe-induced transition of a water molecule from molecular to hydroxyl pair form.
\item Movie S4: Simulation of a probe-induced orientation change of a water molecule in hydroxyl pair form. 
\end{itemize}

\bmhead{Acknowledgments}

This work was supported by NIMS grants (AG2030 and AM2100), by several Grant-in-Aid for Scientific Research (19H05789, 21H01812, 21K18876, 22H00285, 23KJ1516, 24K01350) from the Ministry of Education, Culture, Sports, Science and Technology of Japan (MEXT) and by the Spanish Ministry of Science, Innovation and Universities (MCIU), through project PID2023--149150OB--I00, and the ``Mar\'{\i}a de Maeztu'' Programme for Units of Excellence in R\&D (CEX2023--001316--M). E.F.V. acknowledges support from the Margarita Salas postdoctoral fellowship (Spanish MIU and European Union – NextGenerationEU) and the MOMENTUM Program (MMT-24-ICP-01, Plan de Recuperaci{\'o}n, Transformaci{\'o}n y Resiliencia – European Union – NextGenerationEU). The Spanish Supercomputing Network (RES) is acknowledged for providing computational resources at the Marenostrum Supercomputer (BSC, Barcelona).

\clearpage
\newpage

\bibliography{CeriaBiblio}

\begin{thebibliography}{10}
\bibitem{note0}
\bibinfo{note}{This duality originates from the specific lateral distance between the ceria site and its three neighboring oxygen atoms at the CeO$_2$(111) surface and the low energy cost of displacing these oxygen species.}
\bibitem{Note_defects}
\bibinfo{note}{Mostly sub-surface defects characterized by contrast variations in topographic AFM images of the outermost oxygen surface layer. Occasionally, surface oxygen vacancies were also found. The crystals had a back color, so they probably also contained atomic impurities.}
\end{thebibliography}

\end{document}


\title[Article Title]
{
Supplementary Information for:\\ Near-surface Defects Break Symmetry in Water Adsorption on CeO$_{2-x}$(111) 
}

\author*[1]{\fnm{Oscar} \sur{Custance}}
\email{custance.oscar@nims.go.jp}
\equalcont{These authors contributed equally to this work.}

\author[2]{\fnm{Manuel} \sur{Gonz\'{a}lez-Lastre}}
\equalcont{These authors contributed equally to this work.}

\author[3]{\fnm{Kyungmin} \sur{Kim}}

\author[2,4]{\fnm{Estefan\'{i}a} \sur{Fernandez-Villanueva}}

\author[2,5]{\fnm{Pablo} \sur{Pou}}

\author[3]{\fnm{Masayuki} \sur{Abe}}

\author[1]{\fnm{Hossein} \sur{Sepehri-Amin}}

\author[1]{\fnm{Shigeki} \sur{Kawai}}

\author[4]{\fnm{M. Ver\'{o}nica} \sur{Ganduglia-Pirovano}}

\author*[2,5]{\fnm{Ruben} \sur{Perez}}
\email{ruben.perez@uam.es}

\affil [1]{\orgname{National Institute for Materials Science (NIMS)}, \orgaddress{\street{1-2-1 Sengen}, \city{Tsukuba}, \postcode{305-0047}, \state{Ibaraki}, \country{Japan}}}

\affil [2]{\orgdiv{Departamento de F\'{i}sica Te\'{o}rica de la Materia Condensada}, \orgname{Universidad Aut\'{o}noma de Madrid}, \orgaddress{ \postcode{28049}, \state{Madrid}, \country{Spain}}}

\affil[3]{\orgdiv{Graduate School of Engineering Science}, \orgname{Osaka University}, \orgaddress{\street{1-3 Machikaneyama}, \city{Toyonaka}, \postcode{560-0043}, \state{Osaka}, \country{Japan}}}

\affil [4]{\orgdiv{Instituto de Cat\'{a}lisis y Petroleoqu\'{i}mica (CSIC)}, \orgaddress{ \postcode{28049}, \state{Madrid}, \country{Spain}}}

\affil [5]{\orgdiv{Condensed Matter Physics Center (IFIMAC)}, \orgname{Universidad Aut\'{o}noma de Madrid}, %
\orgaddress{ \postcode{28049}, \state{Madrid}, \country{Spain}}}

\maketitle

\newpage
\section{Additional experimental information}

\begin{figure}[h!]
\centering
\includegraphics[width=0.99\textwidth]{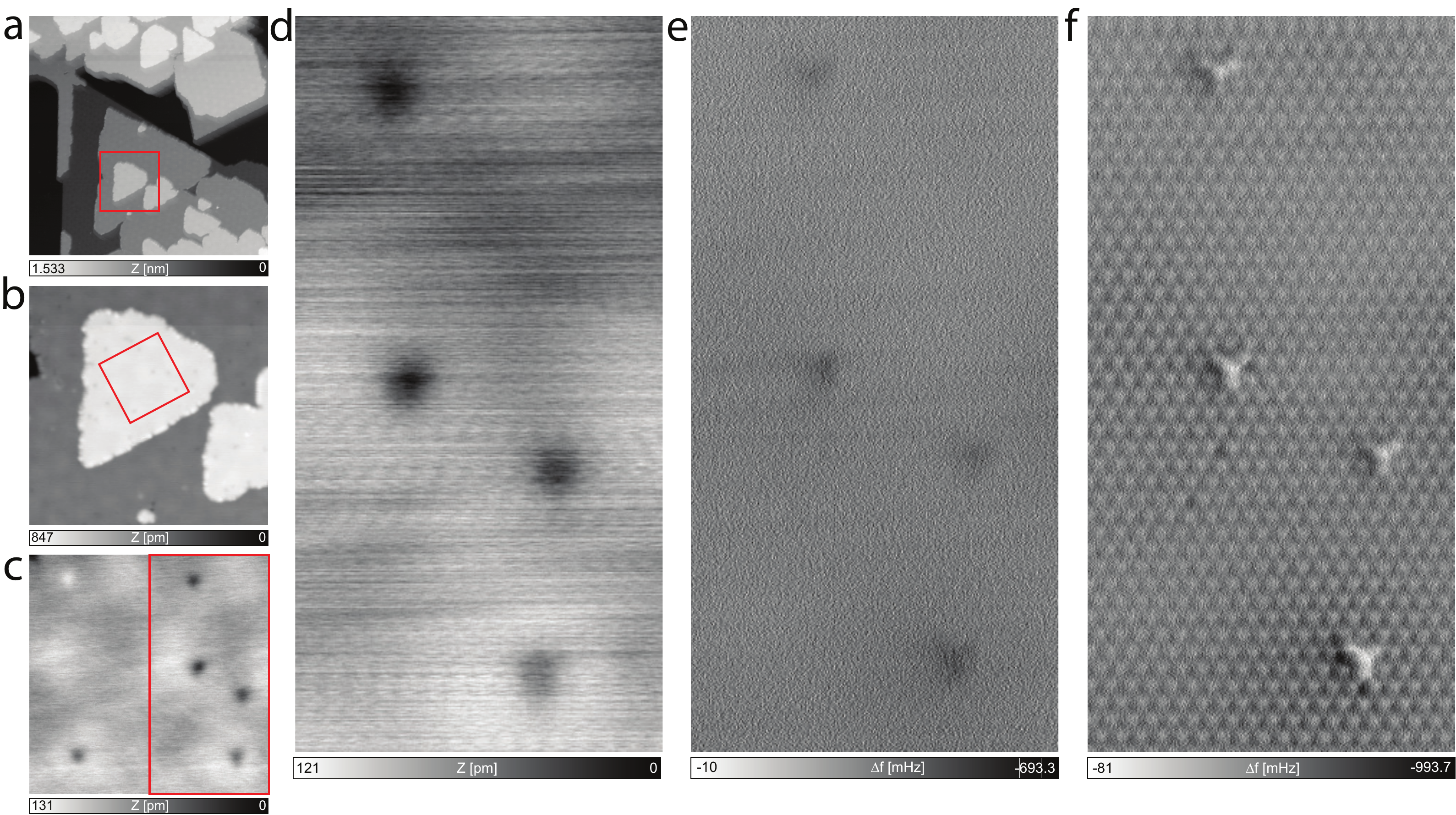}
\caption{ {\bf Surface overview and water molecules location.}
{\bf a}, Topographic STM image of the surface area explored in these experiments. Size is (200$\times$200)~nm${^2}$.   
{\bf b}, Topographic STM image of the CeO$_2$(111) island enclosed by a square in {\bf a}. Size is (50$\times$50)~nm${^2}$.
{\bf c}, Topographic STM image of the surface terraces highlighted by a rectangle in {\bf b}. Size is (15$\times$15)~nm${^2}$.
{\bf d}, Topographic STM image of the surface area displayed in Fig.~1a.
{\bf e}, Constant-height AFM image at the probe-surface separation defined by the topographic set point in {\bf d}.
{\bf f}, AFM image displayed in Fig.~1a.
In these experiments, we used a sample bias voltage ($V$) of 2.5~V and tunneling current set point ($I_{set}$) of 2~pA for STM topographic imaging, an oscillation amplitude ($A$) of 60~pm and the sensor free oscillation resonant frequency ($f_0$) was 994230 Hz. 
}
\label{FigS1}
\end{figure}

\clearpage

\begin{figure}[p!]
\centering
\includegraphics[width=0.99\textwidth]{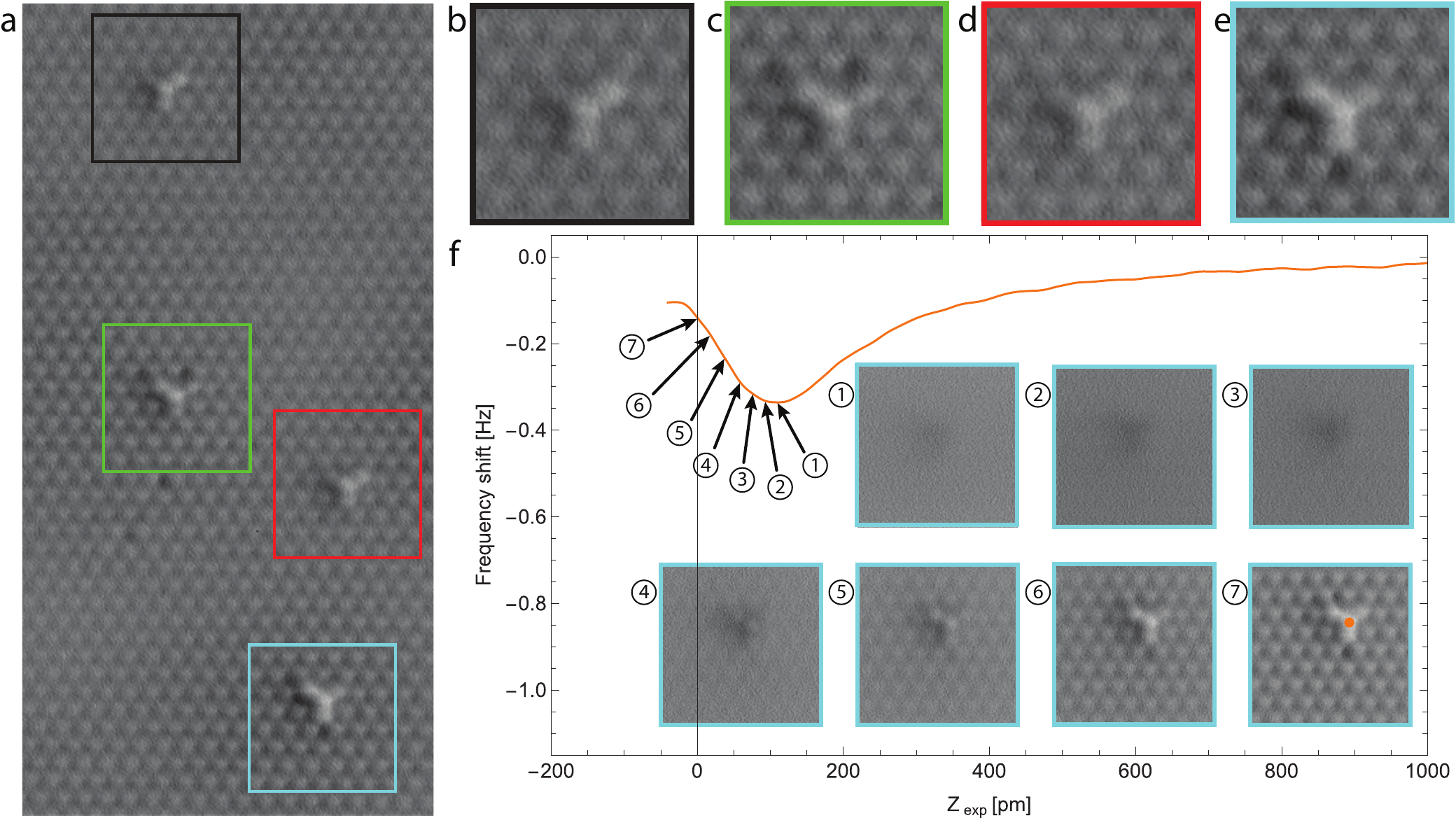}
\caption{{\bf AFM contrast of the water molecule.}
{\bf a},  Constant-height AFM image acquired several hours after the one displayed in Fig.~1a, in the same location and with the same acquisition parameters.
The four water molecules appearing in the image have been enclosed by squares to label each molecule.
{\bf b} to {\bf e}, Detailed view of the water molecules appearing in {\bf a}.
The color of the image frame in each panel matches the color of the squares in {\bf a}.
The boomerang-like shape of the molecules appear on the right for  {\bf b} and {\bf d}, and on the left for {\bf c} and {\bf e}.
Notice that the boomerang-like shape orientation of the molecule in {\bf b} changed from being oriented towards the top in Fig.~1a. to being oriented to right in {\bf a}.
These observations exclude a probe artifact as the origin of the boomerang-like shape of the molecules.  
{\bf f}, Graph showing the AFM signal with the probe-surface distance measured at the center of the water molecule displayed in {\bf e}.
The inset shows the variation of the AFM contrast over the water molecule in {\bf e} with the probe-surface separation. 
{\bf 1} to {\bf 7} present constant-height AFM images acquired successively approaching the probe towards the surface from a distance at which some AFM signal over the molecule is detected ({\bf 1}) to the separation at which the image in {\bf a} was obtained  ({\bf 7}).
The point in the force spectroscopy curve at which each image was measured is indicated by arrows labelled with the same number as the images.  
}
\label{FigS2}
\end{figure}

\clearpage

\begin{figure}[p!]
\centering
\includegraphics[width=0.99\textwidth]{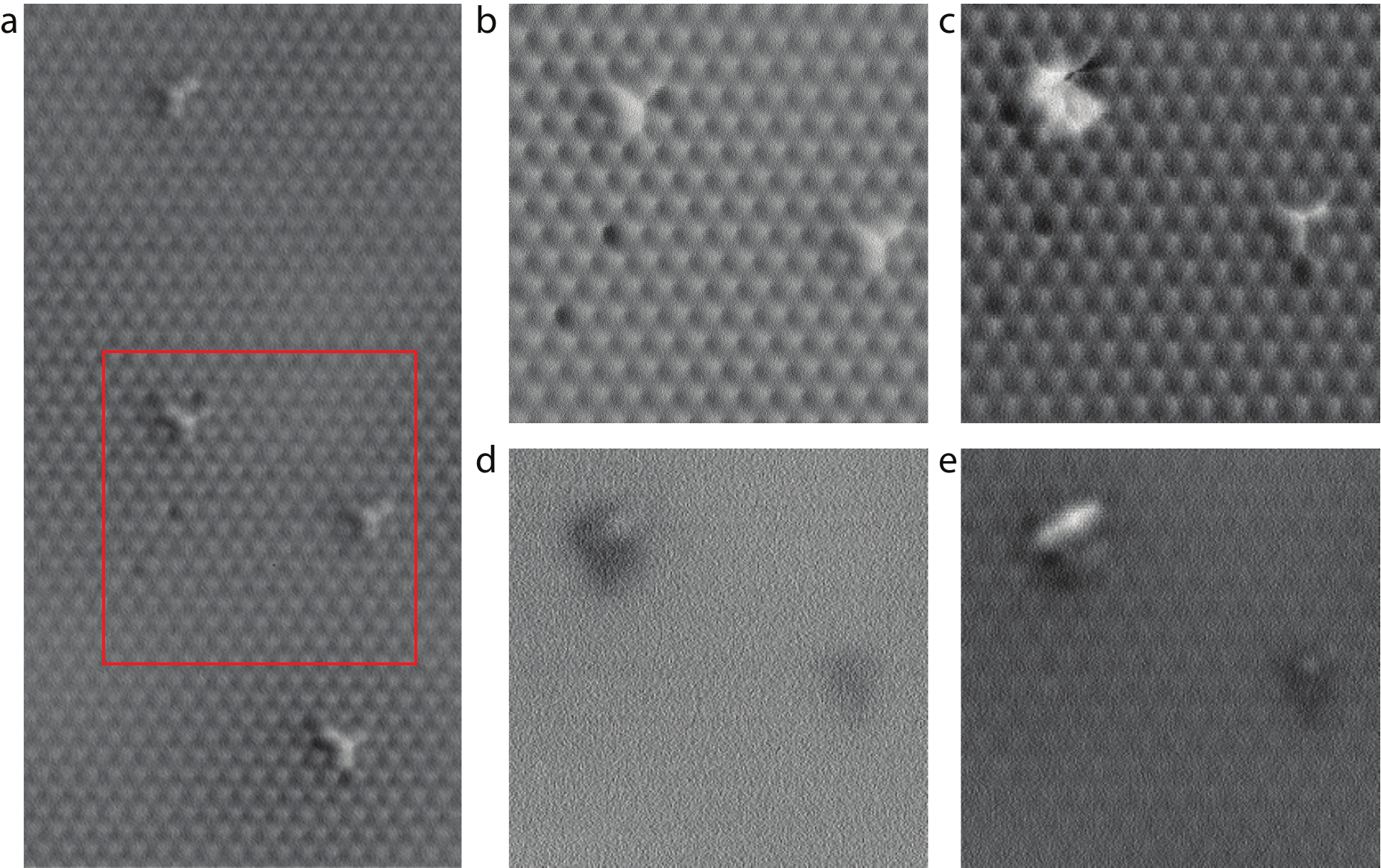}
\caption{{\bf Water lateral manipulation attempt.}
{\bf a}, AFM image displayed in Fig.~S2a.
{\bf b}, Constant-height AFM image of the region highlighted in {\bf a}.
After taking this image we attempted the lateral manipulation of the water molecule shown in the upper left part of the image to a neighboring Ce site following the procedure we reported in {\it Nat. Mater.} {\bf 4}, 156 (2005) (DOI: 10.1038/nmat1297) that successfully enabled us to manipulate individual water molecules on the (111) surface of CeO${_2}$ single crystals as reported in {\it J. Phys.: Condens. Matter.} {\bf 24}, 084010 (2012) (DOI: 10.1088/0953-8984/24/8/084010).
However, we were unable to induce the jump of the water molecule even considerably increasing the probe-surface interaction.
Instead of manipulating the water molecule, we induced the modification of the probe-surface interface.
{\bf c}, Constant-height AFM image of the region highlighted in {\bf a} after the manipulation attempt acquired at the same height as for {\bf b}.
{\bf d} and {\bf e}, Constant-height AFM image of the region highlighted in {\bf a} retracting the probe by 173~pm and 114~pm, respectively, from the height the image in {\bf c} was measured.  
}
\label{FigS3}
\end{figure}

 \begin{figure}[p!]
\centering
\includegraphics[width=0.99\textwidth]{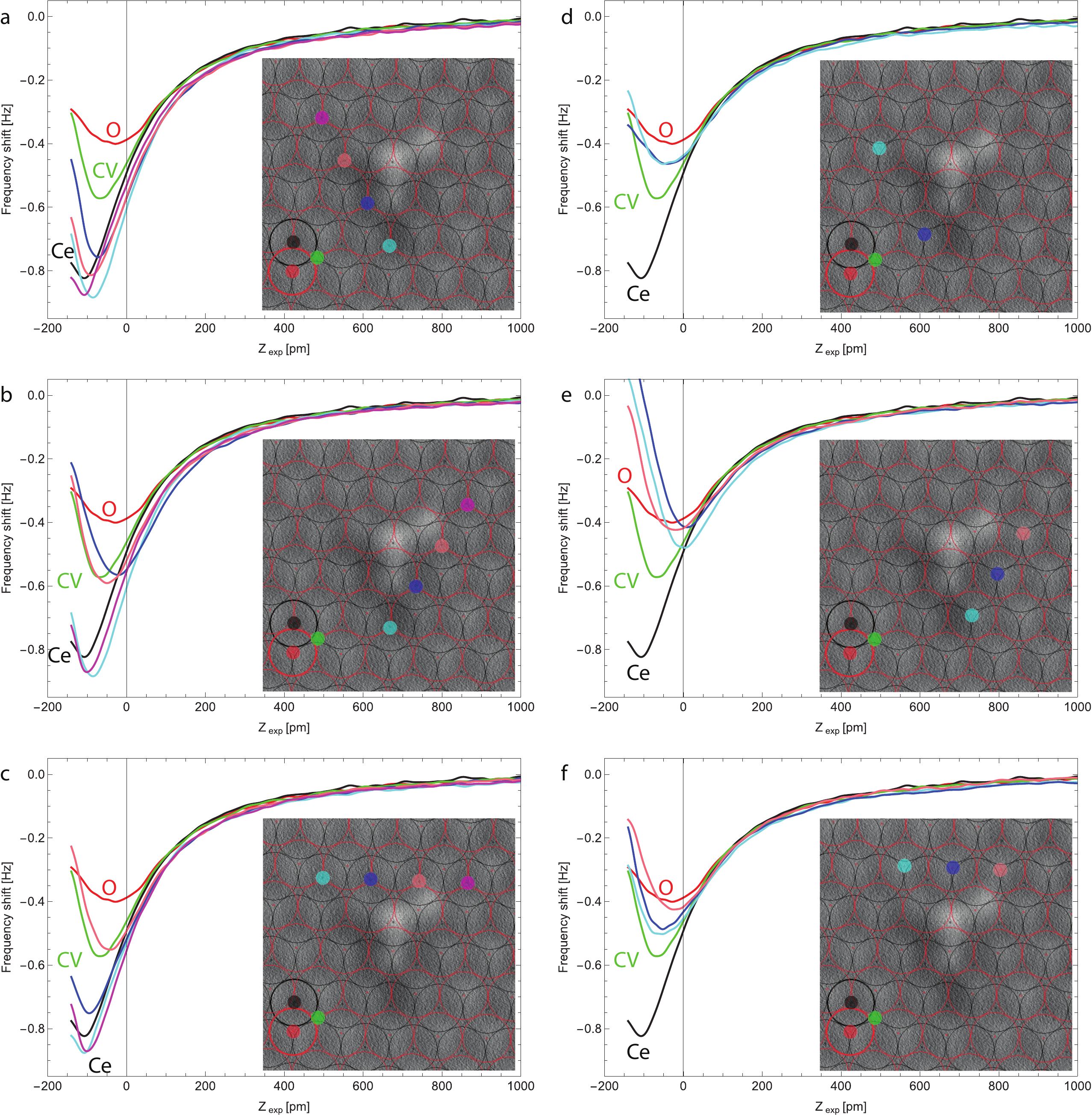}
\caption{
{\bf Analysis of the force spectroscopy measurements displayed in Fig.~7.}
{\bf a} AFM signal over the cerium positions on the left side respect to the water molecule. 
{\bf b} AFM signal over the cerium positions on the right side respect to the water molecule. 
{\bf c} AFM signal over the cerium positions on the top respect to the water molecule. 
{\bf d} AFM signal over the oxygen positions on the left side respect to the water molecule. 
{\bf e} AFM signal over the oxygen positions on the right side respect to the water molecule. 
{\bf f} AFM signal over the oxygen positions on the top respect to the water molecule.
In all the panels, curves measured over the three surface sites of the CeO${_2}$(111) surface (Ce in black, O in red and coordination vacancy in green) far from the water molecule are included for comparison.
Color of the curves matches the color of the circles indicating the acquisition position in the inset image.
A constant height AFM image measured before the spectroscopic acquisition is included in each panel.
The black and red lattices superimposed to the image highlight the cerium and oxygen surface positions, respectively.
The origin in the distance axis indicates the probe-surface separation the image was acquired. 
}
\label{FigS11}
\end{figure}
\clearpage
\section{Additional information for DFT calculations on water adsorption and dissociation}

\subsection{Calculations on the 3$\times$3 unit cell}

As discussed in the main text, the existence of a SSO$_V$ and the corresponding Ce$^{3+}$ changes the surface geometry and breaks its threefold symmetry, resulting in unequivalent sites for H$_2$O adsorption. In agreement with previous studies 
(Ref.~30 in the main text),
the water molecule rarely stabilizes forming two hydrogen bonds with the surface oxygen atoms (structures {\bf9}, {\bf14} and {\bf18} in Fig.~\ref{FigS5}). Instead, it prefers to form one hydrogen bond with an oxygen atom of the surface while the second hydrogen atom moves upwards, forming an angle of $\sim$120$^o$ with the surface 
(e.g. structure {\bf3} in Fig.~\ref{FigS5})
Since each cerium atom is surrounded by three surface oxygen atoms, the water molecule can form up to three different hydrogen bonds, and given that the free hydrogen atom can also be oriented to either side of this hydrogen bond  
(e.g. see structures {\bf10} and {\bf11} in Fig.~\ref{FigS5}),
ultimately there are up to six different adsorption structure possibilities for the water molecule per cerium site (Fig.~\ref{FigS5}).

The geometric distortion of the surface due to the SSO$_V$ is stronger for the sites surrounding it and for the Ce$^{3+}$ sites, which is reflected in a greater adsorption energy change. 
In general, we found that the water molecule feels repulsion from the surface oxygen atoms, so that the closer they are, the less stable the adsorbed molecule becomes. As a result, adsorption energies correlate with the Ce-O bond distances of the cerium atom the water molecule adsorbs on.

More specifically, in the vicinity of the SSO$_V$ the oxygen atoms move deeper into the surface as well as towards the vacancy, which causes the shortening of the two bonds adjacent to the SSO$_V$  ($\sim$2.29~\AA\space vs. 2.37~\AA\space in the fully oxidized surface, see oxygen atom {\bf A} in Fig.~\ref{FigS9}a). 
As a consequence, adsorptions at the Ce$^{4+}$ neighboring the vacancy are the weakest (+0.13 eV), and among the possibilities, the molecule prefers to form the hydrogen bond with the oxygen atom that is furthest from the SSO$_V$. Meanwhile, Ce$^{3+}$ ions are bigger than Ce$^{4+}$, and their Ce-O equilibrium distances are noticeably longer (2.46-2.64~\AA\space Ce$^{3+}$ in green in Fig.~\ref{FigS9}a). This geometric effect compensates for the excess charge of the cerium atom (which in principle should lead to a greater repulsion with the water molecule) and actually leads to the strongest adsorption found.

Naturally, bond changes are amplified for those oxygen atoms that neighbor both the vacancy and a Ce$^{3+}$. In this case, the bigger size of the Ce$^{3+}$ pushes the oxygen atoms even more towards the vacancy, further increasing the shortening effect by -0.04~\AA\space and the elongating effect by 0.14~\AA\space ($\sim$2.25~\AA\space and $\sim$2.64~\AA, respectively, for the oxygen atoms {\bf B} in Fig.~\ref{FigS9}a). Finally, the bond distances and energies obtained on the Ce$^{4+}$ far from the vacancy are similar to the ones of the fully oxidized CeO${_2}$(111) surface, with the energy progressively getting less favorable as the cerium site approaches the location of the vacancy (Fig.~\ref{FigS6}c-e and Fig.~\ref{FigS9}a).

All in all, the adsorption energies and Ce-O bond distances of the different cerium atom sites follow the trend (Figs.~\ref{FigS9}a~and~\ref{FigS5}): 
Ce$^{3+}$ ($\Delta$E$_{ads}$$\approx$~-0.79~eV, r(Ce-O)$\approx$~2.46~to~2.64~\AA) $>$ 
Ce$^{4+}$ far from SSO$_V$ ($\Delta$E$_{ads}$$\approx$~-0.72 eV, r(Ce-O)$\approx$~2.30~to~2.50~\AA) $>$ 
Ce$^{4+}$ close to SSO$_V$ ($\Delta$E$_{ads}$$\approx$~-0.66 eV, r(Ce-O)$\approx$~2.25~to~2.27~\AA). 

\subsection{Calculations on the 4$\times$4 unit cell}

To ensure that our results on the stability of water on the Ce$^{3+}$, either in molecular or dissociated species (Fig.~\ref{FigS7}), were not significantly affected by the two Ce$^{3+}$ atoms being neighbors on the (3$\times$3) slab, we also calculated the crucial structures on the reduced CeO${_2}$(111) surface using a (4$\times$4) model (Figs.~\ref{FigS7}a~and~\ref{FigS8}b).
Then, we also tested whether these results would qualitatively hold when the two Ce$^{3+}$ atoms produced by the SSO$_V$ are in the second energetically most stable configuration reported instead (see Ref.~61 in the main text), which is that of Ce$^{3+}$ atoms labelled {\bf11} and {\bf 13} in Fig.~\ref{FigS9}d (Figs.~\ref{FigS7}b~and~c~and~\ref{FigS8}c~and~d, respectively).

\begin{figure}[p!]
\centering
\includegraphics[width=0.78\textwidth]{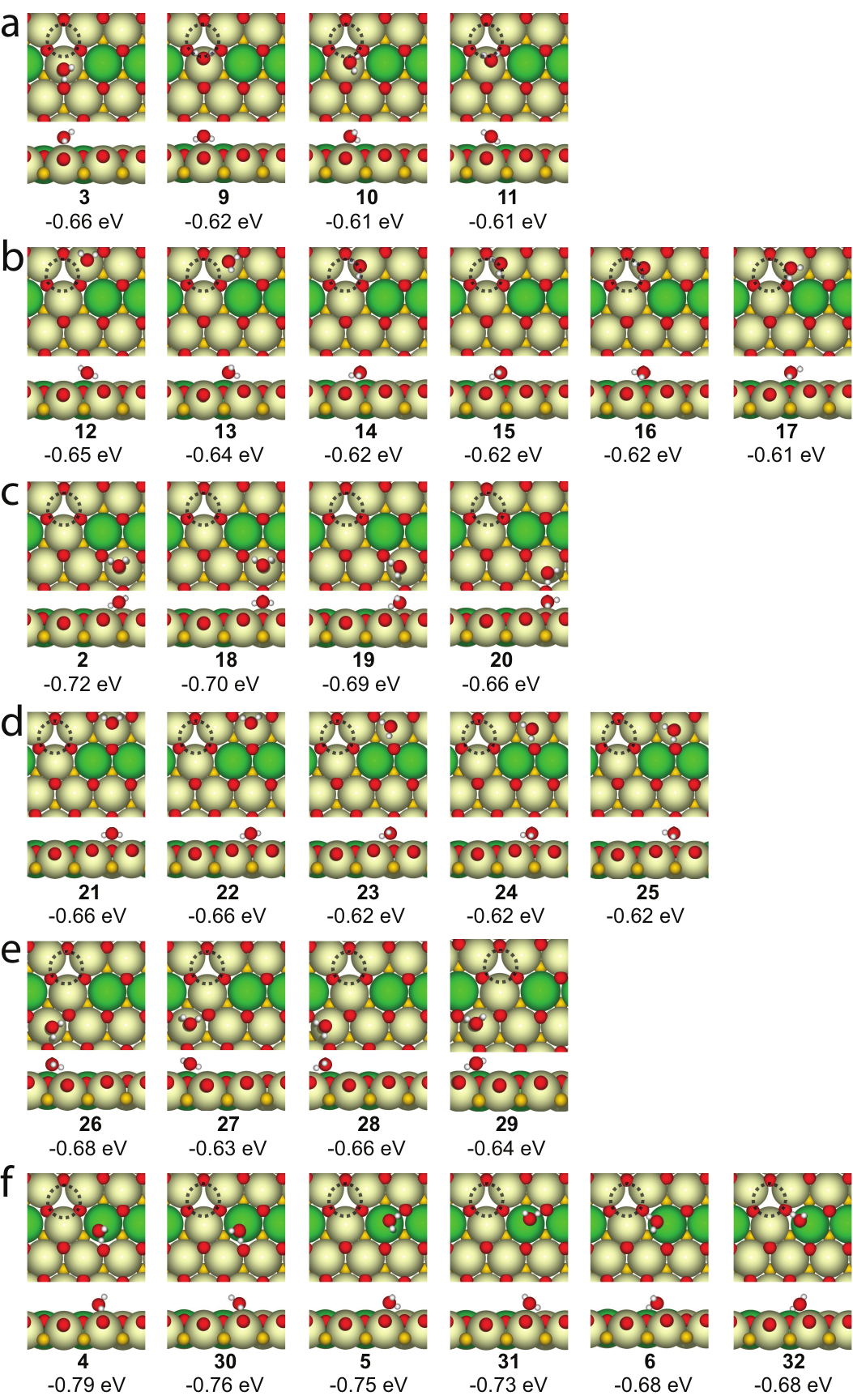}
\caption{{\bf DFT optimized water adsorption structures in the presence of a subsurface oxygen vacancy using a (3$\times$3)
unit cell}.
Calculated structures and energies for the adsorption of a water molecule at inequivalent cerium sites of the CeO$_2$(111) surface. 
The energetically most stable configuration for the position of the Ce$^{3+}$ atoms generated by the subsurface oxygen vacancy was considered (see Ref.~61 in the main text).
{\bf a} and {\bf b}, Adsorption on inequivalent Ce$^{4+}$ atoms close to the vacancy.
{\bf c} to {\bf e}, Adsorption on inequivalent Ce$^{4+}$ far from the vacancy. 
{\bf f}, Adsorption on one of the two equivalent Ce$^{3+}$ atoms of the structure. 
The energies with respect to the slab + water molecule are indicated under each structure.
The position of the subsurface oxygen vacancy is highlighted by a dotted circumference. 
Ce$^{3+}$, Ce$^{4+}$, surface oxygen atoms and subsurface oxygen atoms are represented in green, beige, red and yellow, respectively.
}
\label{FigS5}
\end{figure}

\begin{figure}[p!]
\centering
\includegraphics[width=0.99\textwidth]{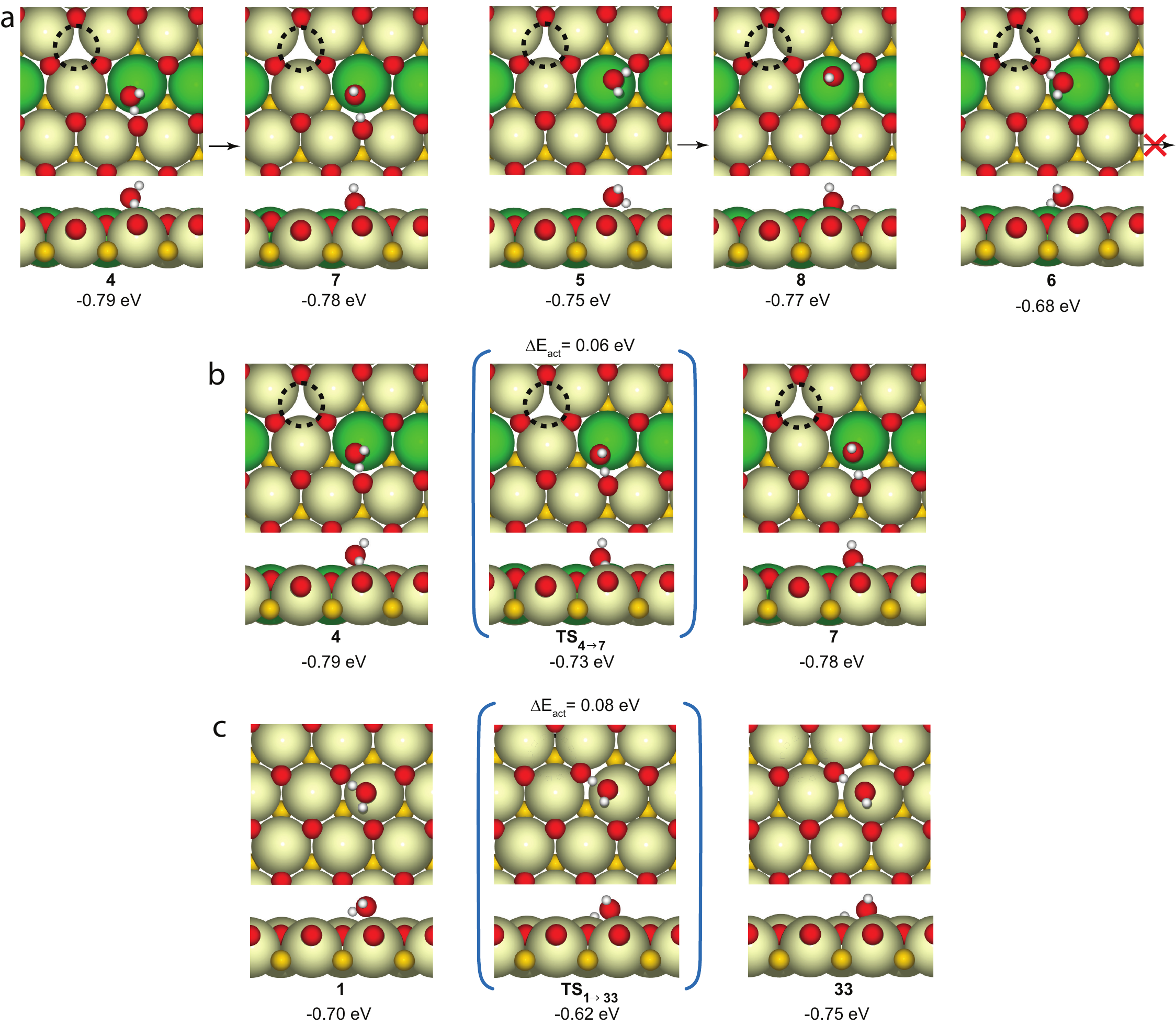}
\caption{{\bf DFT calculated dissociation and transition states of the water molecule using a (3$\times$3) 
unit cell.}
{\bf a}, 
Molecular and dissociated
water structures (H$_2$O~$\rightarrow$~OH + H) on the reduced CeO$_2$(111) surface from Fig.~2 of the main text.
{\bf b} and to {\bf c}, Water dissociation on the reduced and fully oxidized surface, respectively.
The corresponding transition states ({\bf TS}) structures and activation energies are shown within parenthesis.
The energies with respect to the slab + water molecule are indicated under each structure.
 The position of the subsurface oxygen vacancy is highlighted by a dotted circumference. 
Ce$^{3+}$, Ce$^{4+}$, surface oxygen atoms and subsurface oxygen atoms are represented in green, beige, red and yellow, respectively.
}
\label{FigS6}
\end{figure}

\begin{figure}[p!]
\centering
\includegraphics[width=0.85\textwidth]{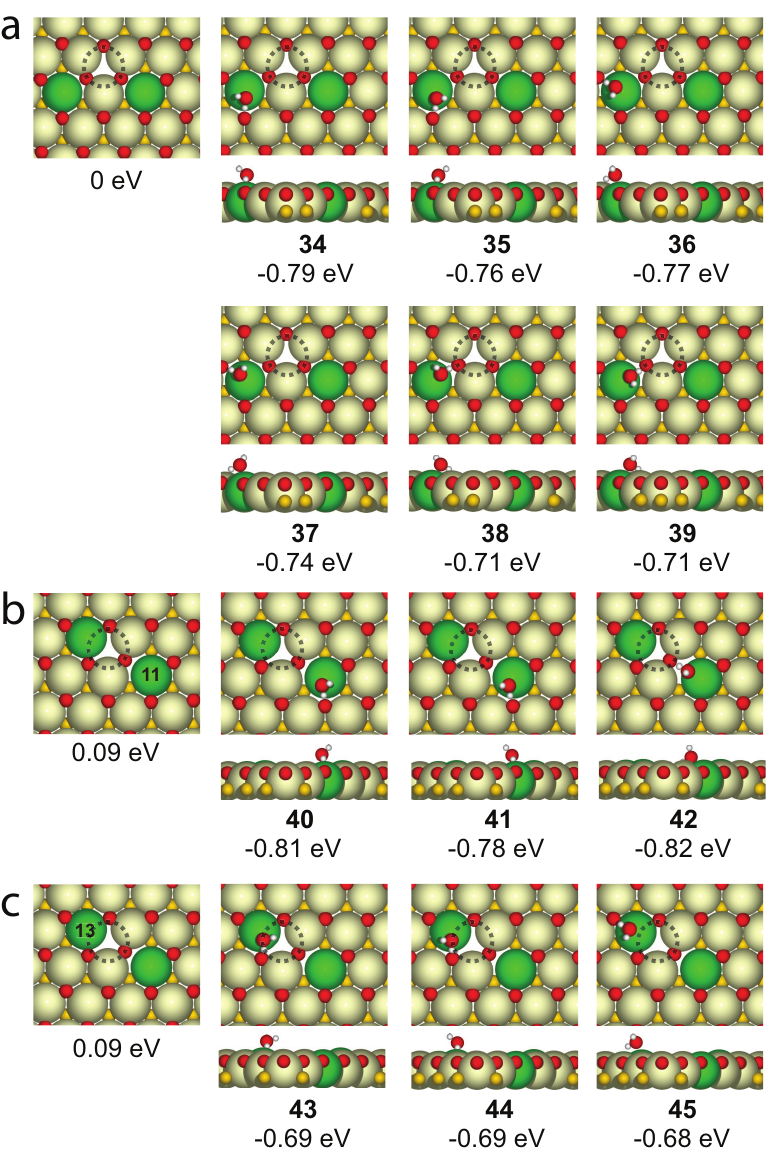}
\caption{{\bf DFT optimized structures for the water adsorption on nonequivalent Ce$^{3+}$ atoms using a (4$\times$4)
unit cell.
}
Structures for the water adsorption on each nonequivalent Ce$^{3+}$ considering the most stable and second most stable positions for the the Ce$^{3+}$ atoms generated by the presence of a subsurface oxygen vacancy on the CeO$_2$(111) surface.
{\bf a}, Considering the energetically most stable configuration with the two Ce$^{3+}$ atoms in equivalent positions.
{\bf b} and {\bf c}, Considering the energetically second most stable configuration with the two Ce$^{3+}$ in nonequivalent positions.
These Ce$^{3+}$ atoms are labelled as {\bf 11} and {\bf 13} for reference. 
The energies with respect to the slab + water molecule are indicated under each structure.
 The position of the subsurface oxygen vacancy is highlighted by a dotted circumference. 
Ce$^{3+}$, Ce$^{4+}$, surface oxygen atoms and subsurface oxygen atoms are represented in green, beige, red and yellow, respectively.
}
\label{FigS7}
\end{figure}

\begin{figure}[p!]
\centering
\includegraphics[width=0.87\textwidth]{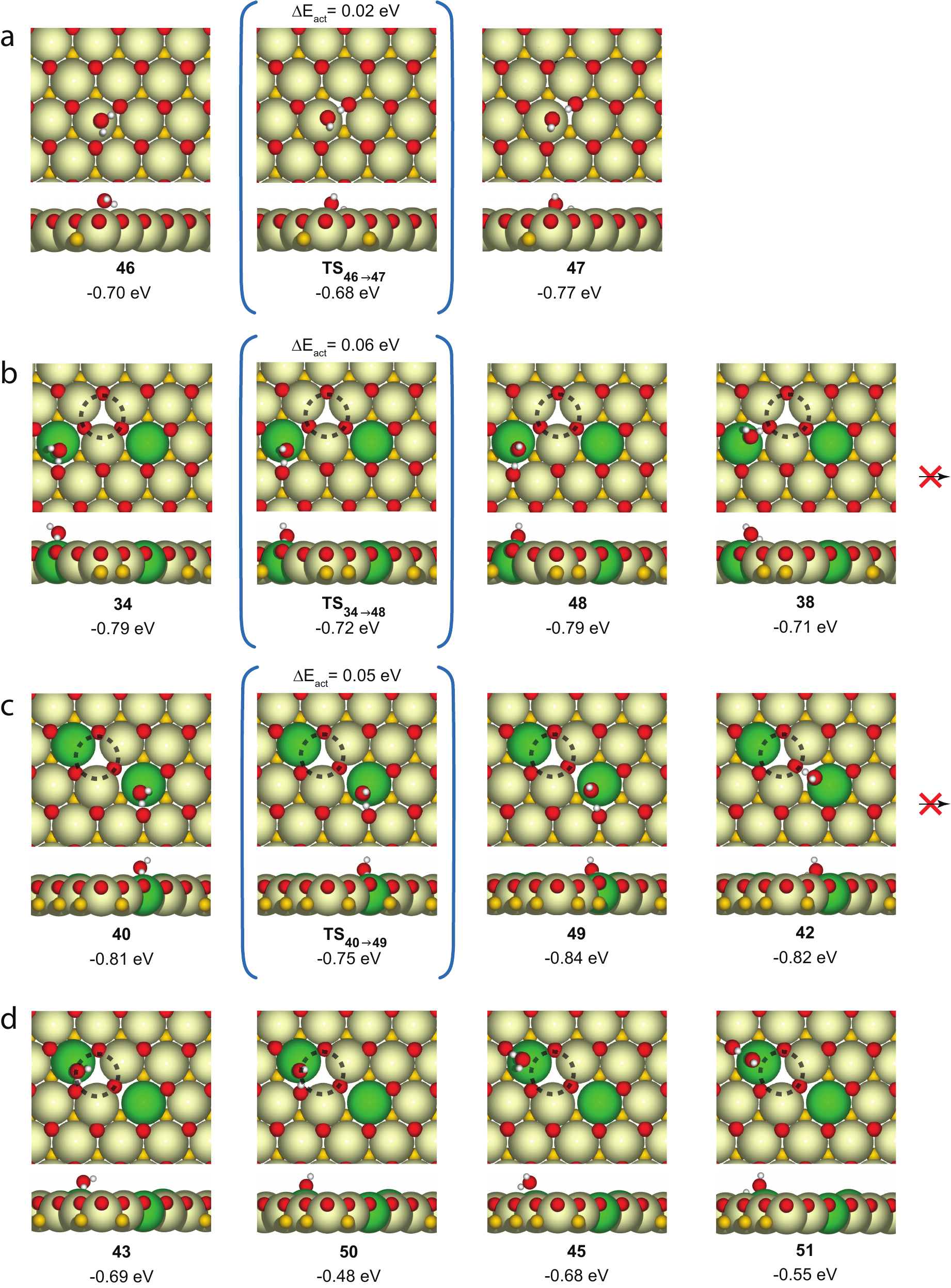}
\caption{{\bf DFT calculated dissociation and transition states of the water molecule using a (4$\times$4)
unit cell
.} 
Dissociated water (H$_2$O~$\rightarrow$~OH + H) and transition states ({\bf TS}) structures and activation energies for:
{\bf a}, a fully oxidized CeO$_2$(111) surface;
{\bf b} a reduced surface considering the energetically most stable configuration for the position of the Ce$^{3+}$ atoms;
and {\bf c} and {\bf d}, a reduced surface considering the  second energetically most stable configuration for the position of the Ce$^{3+}$ atoms, labelled as {\bf 11} and {\bf 13} in Figs.~S14d~and~S7b~and~c. 
No OH + H stable configurations could be optimized from structures {\bf 38} and {\bf 42} from Fig.S7. 
The corresponding transition states ({\bf TS}) structures and activation energies are shown within parenthesis.
No TS were calculated for {\bf d} since the dissociated structures were significantly less stable than the molecular ones.
The energies with respect to the slab + water molecule are indicated under each structure.
The position of the subsurface oxygen vacancy is highlighted by a dotted circumference. 
Ce$^{3+}$, Ce$^{4+}$, surface oxygen atoms and subsurface oxygen atoms are represented in green, beige, red and yellow, respectively.
}
\label{FigS8}
\end{figure}

We found no qualitative differences employing the former: the six water adsorption configurations follow the same trend in relative energies as for the calculations using the (3$\times$3) slab (Fig.~\ref{FigS5}f and~\ref{FigS7}a), and again the OH + H dissociated structure for the least stable configuration cannot be stabilized ({\bf 38} in Fig.~\ref{FigS8}).
With the (4$\times$4) surface, the Ce$^{3+}$ atoms have more space to expand, which is reflected in small variations of the Ce-O distances with respect to the (3$\times$3) slab:  2.672\AA, 2.494\AA\space and 2.512\AA\space for the (4$\times$4), and 2.642\AA, 2.457\AA\space and 2.537\AA\space for the (3$\times$3)(Figs.~\ref{FigS9}a~and~c). 
Nevertheless, the adsorption energies for the water molecule only vary within 0.03~eV (Fig.~\ref{FigS7}).

Regarding the second most stable (4$\times$4) model, note that the Ce$^{3+}$ atom labelled {\bf11} is in the same position with respect to the vacancy as the Ce$^{3+}$ atoms previously discussed, but the new location of the second Ce$^{3+}$ atom (labeled {\bf13}) causes large modifications to the Ce-O bond distances (Fig.~\ref{FigS9}d).
The more dramatic change is for oxygen atom {\bf B$^\prime$} in Fig.~\ref{FigS9}d, which moves towards the SSO$_V$ to a great extent, leading to a Ce-O bond that is substantially longer (2.854\AA\space vs 2.671\AA, Fig.~\ref{FigS9}c~and~d). As a consequence, adsorption energies on Ce$^{3+}$ {\bf11} are even higher than on the other Ce$^{3+}$ configuration (Fig.~\ref{FigS7}c vs. Fig.~\ref{FigS7}a). Also, note that the adsorption energy for a water molecule forming a hydrogen bond with the surface oxygen atom {\bf B$^\prime$}, which neighbors the SSO$_V$, is no longer the less stable adsorption structure, but the most stable one, due to this dramatic Ce-O bond increase (-0.82~eV for structure {\bf 42} in Fig.~\ref{FigS7}c vs. -0.71~eV for structures {\bf 38} and {\bf 39} in Fig.~\ref{FigS7}a). In spite of this, we were still unable to generate a stable dissociated OH + H structure from structure {\bf42} in  Fig.~\ref{FigS7}c, and thus the Ce$^{3+}$ {\bf11} site remains qualitatively similar to the Ce$^{3+}$ in the previous configurations (Fig.~\ref{FigS9}a~and~c).

Regarding the Ce$^{3+}$ atom labeled {\bf13}, which neighbors the SSO$_V$, we found that its Ce-O bond distances are shorter than for any of the other Ce$^{3+}$ atoms presented in this work (2.36-2.46\AA\space, Fig.~\ref{FigS9}d). Indeed, these bond distances are the result of a compromise between the bigger size of the Ce$^{3+}$, which is more stable with larger Ce-O distances, and the oxygen atoms {\bf A$^\prime$}, that are more stable towards the vacancy, i.e. with smaller Ce-O distances. As a result of these constrained Ce-O bond distances for a Ce$^{3+}$, adsorption energies are lower (about -0.69 eV).
Finally, we tested several OH + H structures at the Ce$^{3+}$ atom {\bf13}, but they produced even higher energies than for the water adsorption in molecular form, so we concluded that adsorption at this Ce$^{3+}$ would not significantly contribute to the experimental measurements.

\clearpage
\section{Additional information for simulations of AFM images and spectroscopy}

\begin{figure}[h!]
\centering
\includegraphics[width=0.50\textwidth]{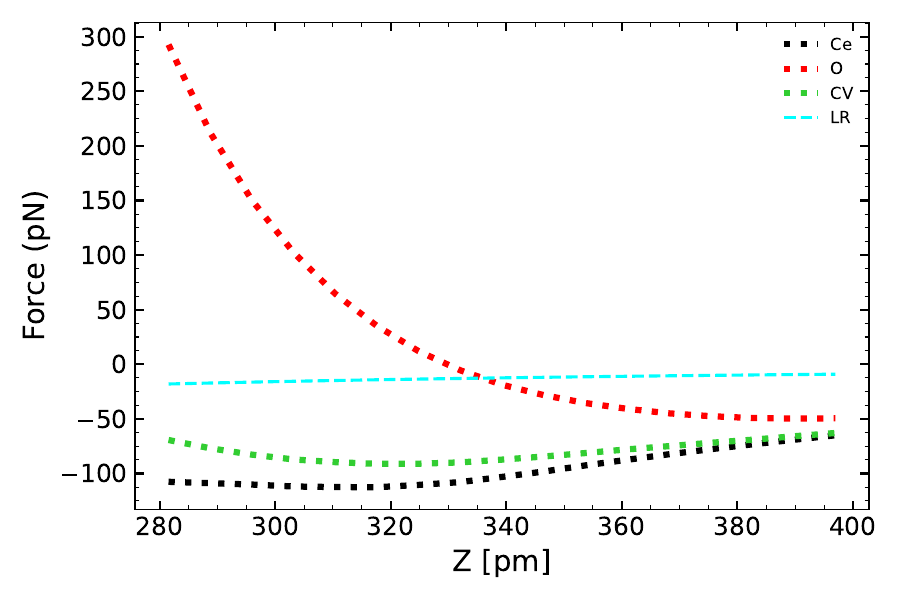}\includegraphics[width=0.50\textwidth]{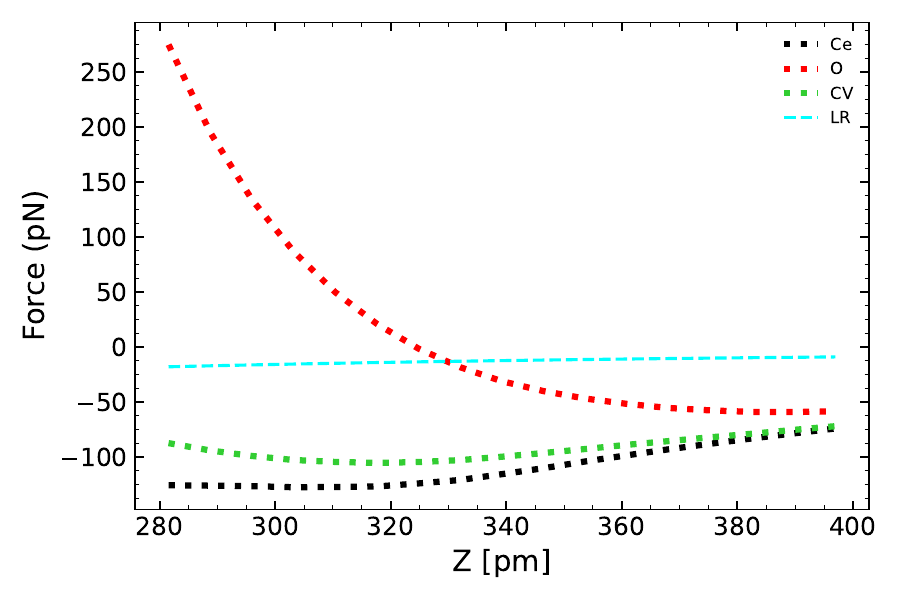}
\caption{{\bf Calculated force curves over a O (red), a Ce (black) and a CV (green) sites on a fully oxidized CeO$_{2}$(111) with our model probe}. Our model for the probe combines DFT forces with a CO molecule (left panel) with a long-range contribution (cyan dashed curve), representing the macroscopic part of the probe, fitted to the experimental results at large distances and added to the DFT forces (right panel). These total forces are the ones used to calculate the frequency shift curves shown in Fig.~1 in the main text.} %
\label{FigSX_forces_Fig01}
\end{figure}

\begin{figure}[h!]
\centering
\includegraphics[width=0.99\textwidth]{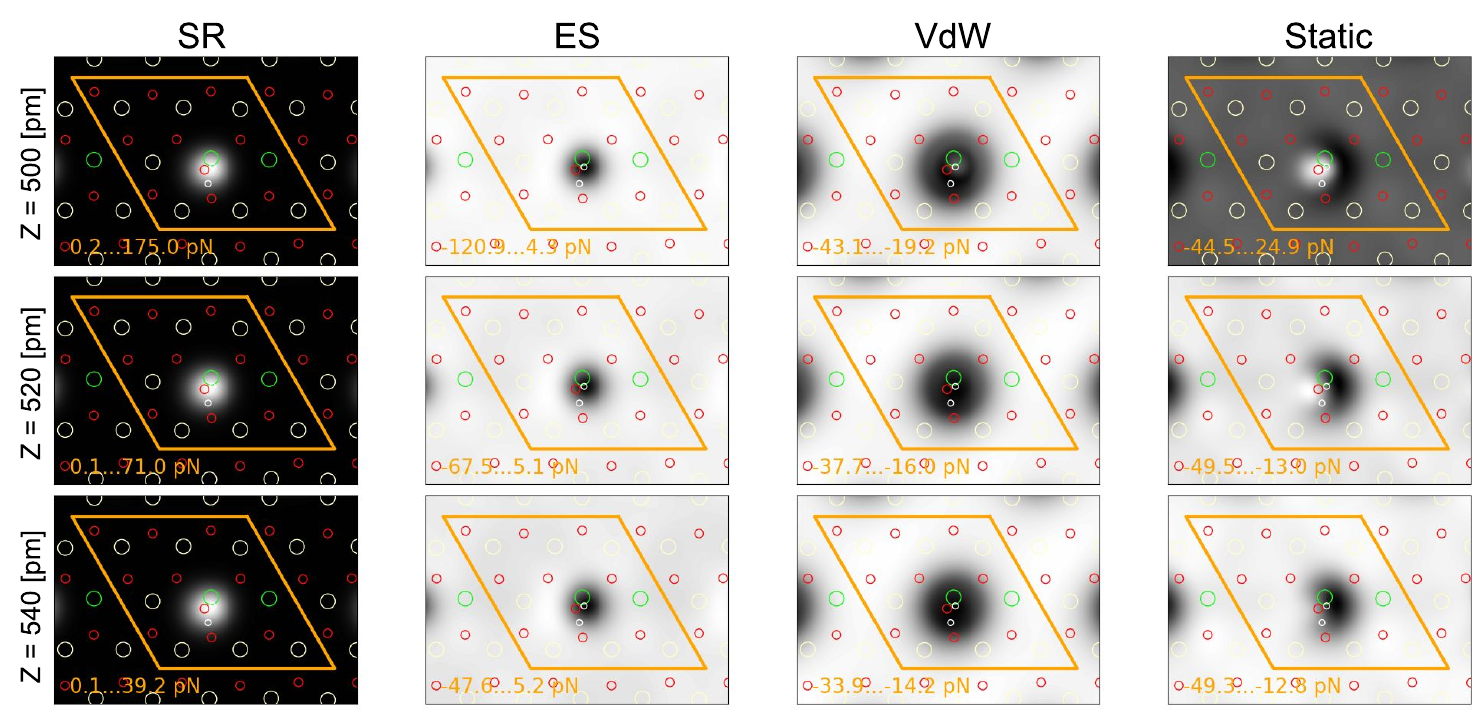}
\caption{{\bf FDBM force decomposition for a water molecule adsorbed in molecular form over a Ce$^{3+}$.}
Decomposition of the AFM contrast obtained from Full Density-Based Model (FDBM) calculations for the water molecule in its molecular configuration adsorbed on a Ce$^{3+}$ site of the reduced CeO$_2$(111) surface (structure {\bf 4} in Fig.~2c).
The total AFM signal (static) is split into short-range Pauli repulsion (SR), electrostatic (ES), and van der Waals (vdW) components.}
\label{FigS12}
\end{figure}

\begin{figure}[h!]
\centering
\includegraphics[width=0.99\textwidth]{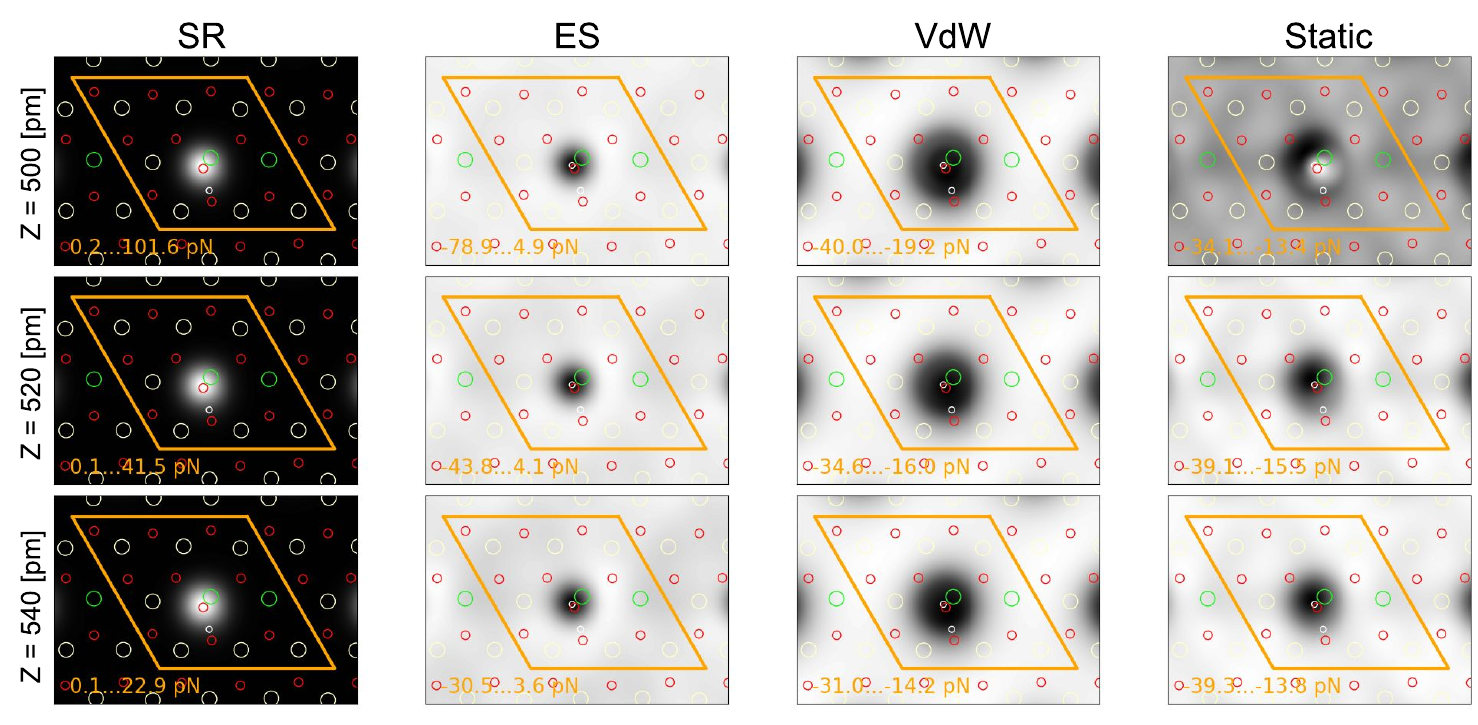}
\caption{{\bf FDBM force decomposition for a water molecule adsorbed in dissociated form 
(as a hydroxyl pair)
over a Ce$^{3+}$.}
Decomposition of the AFM contrast obtained from Full Density-Based Model (FDBM) calculations for the water molecule in its dissociated configuration (OH + H) adsorbed on a Ce$^{3+}$ site of the reduced CeO$_2$(111) surface (structure {\bf 7} in Fig.~2d).}
\label{FigS13}
\end{figure}

\begin{figure}[h!]
\centering
\includegraphics[width=0.50\textwidth]{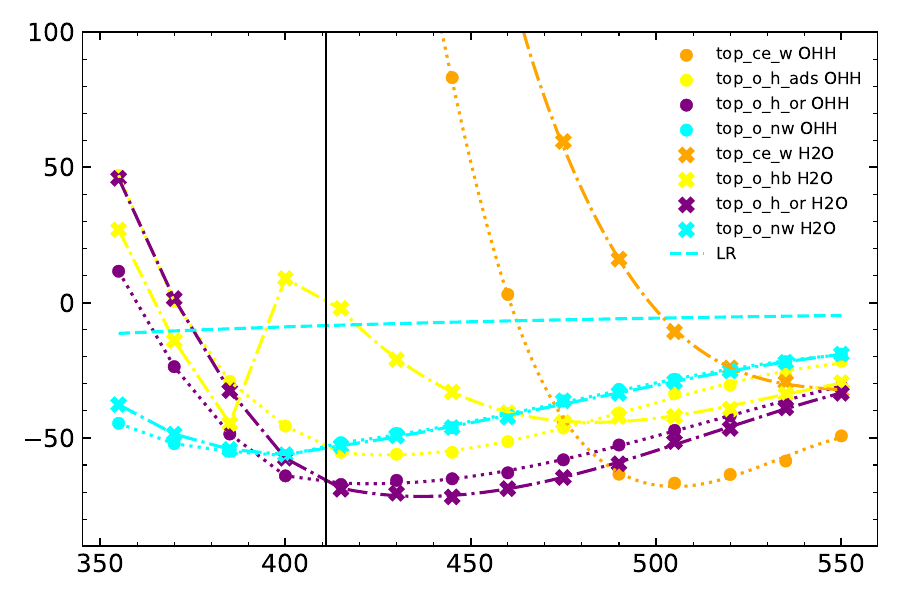}\includegraphics[width=0.50\textwidth]{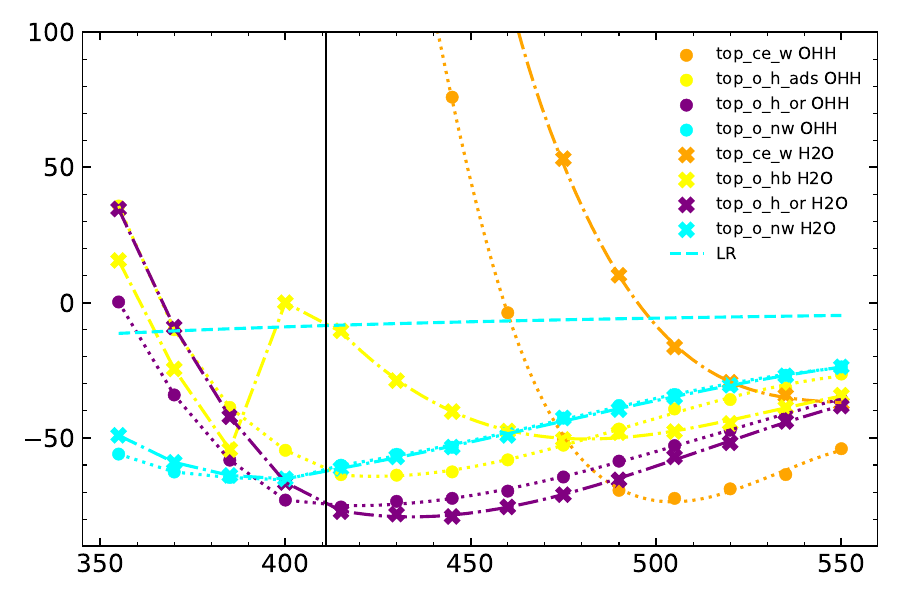}
\caption{{\bf Calculated force curves on different sites for the two possible adsorption states of water, molecular and hydroxil pair, on a fully oxidized CeO$_{2}$(111) with our model probe}. Our model for the probe combines DFT forces with a CO molecule (left panel) with a long-range contribution (cyan dashed curve), representing the macroscopic part of the probe, fitted to the experimental results at large distances and added to the DFT forces (right panel). These total forces are the ones used to calculate the frequency shift curves shown in Fig.~4 in the main text. We use the same color code to  identify the sites}
\label{FigSX_forces_Fig04}
\end{figure}

\begin{figure}[h!]
\centering
\includegraphics[width=0.50\textwidth]{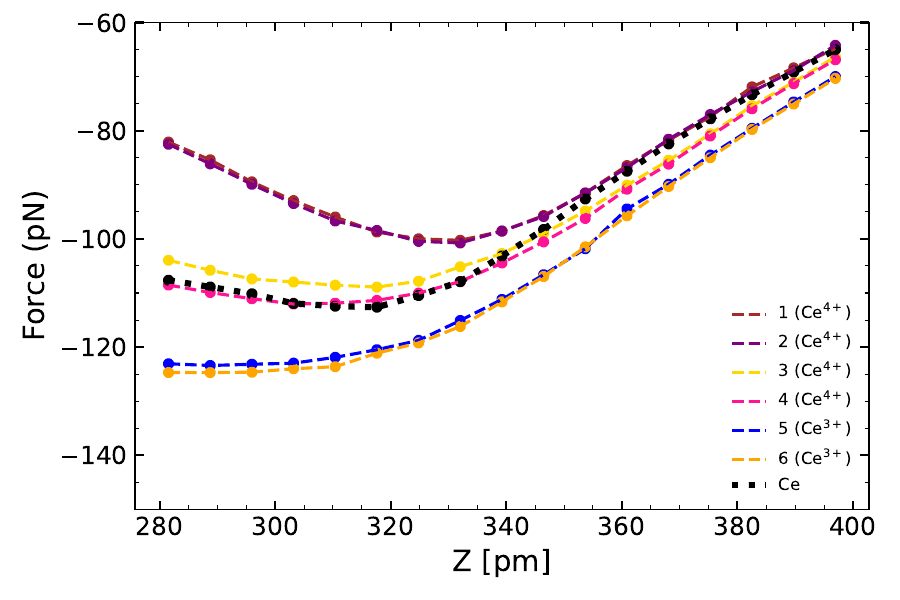}\includegraphics[width=0.50\textwidth]{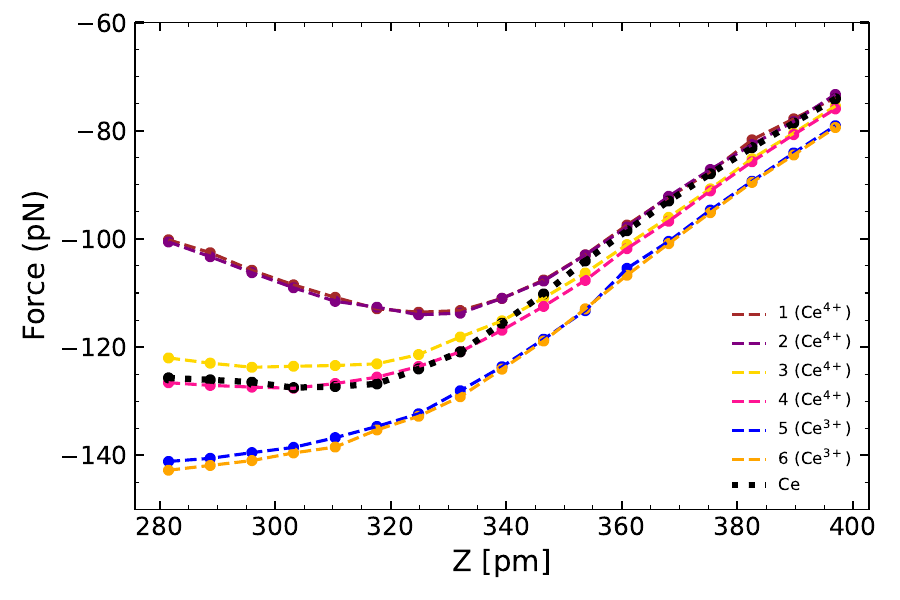}
\caption{{\bf Calculated force curves over different Ce sites close to  a oxygen subsurface vacancy (SSO$_V$).} 
Our model for the probe combines DFT forces with a CO molecule (left panel) with a long-range contribution (not shown in the image), representing the macroscopic part of the probe, fitted to the experimental results at large distances and added to the DFT forces (right panel). These total forces are the ones used to calculate the frequency shift curves shown in Fig.~6 in the main text. We use the same numerical labels to identify the Ce sites:  curves 1-4 correspond to  Ce$^{4+}$ ions and curves 5-6 to the Ce$^{3+}$ ions in the reduced CeO$_{2-x}$(111) surface shown as an inset in Figure 6 of the main text. 
}
\label{FigSX_forces_Fig06}
\end{figure}

\begin{figure}[h!]
\centering
\includegraphics[width=0.99\textwidth]{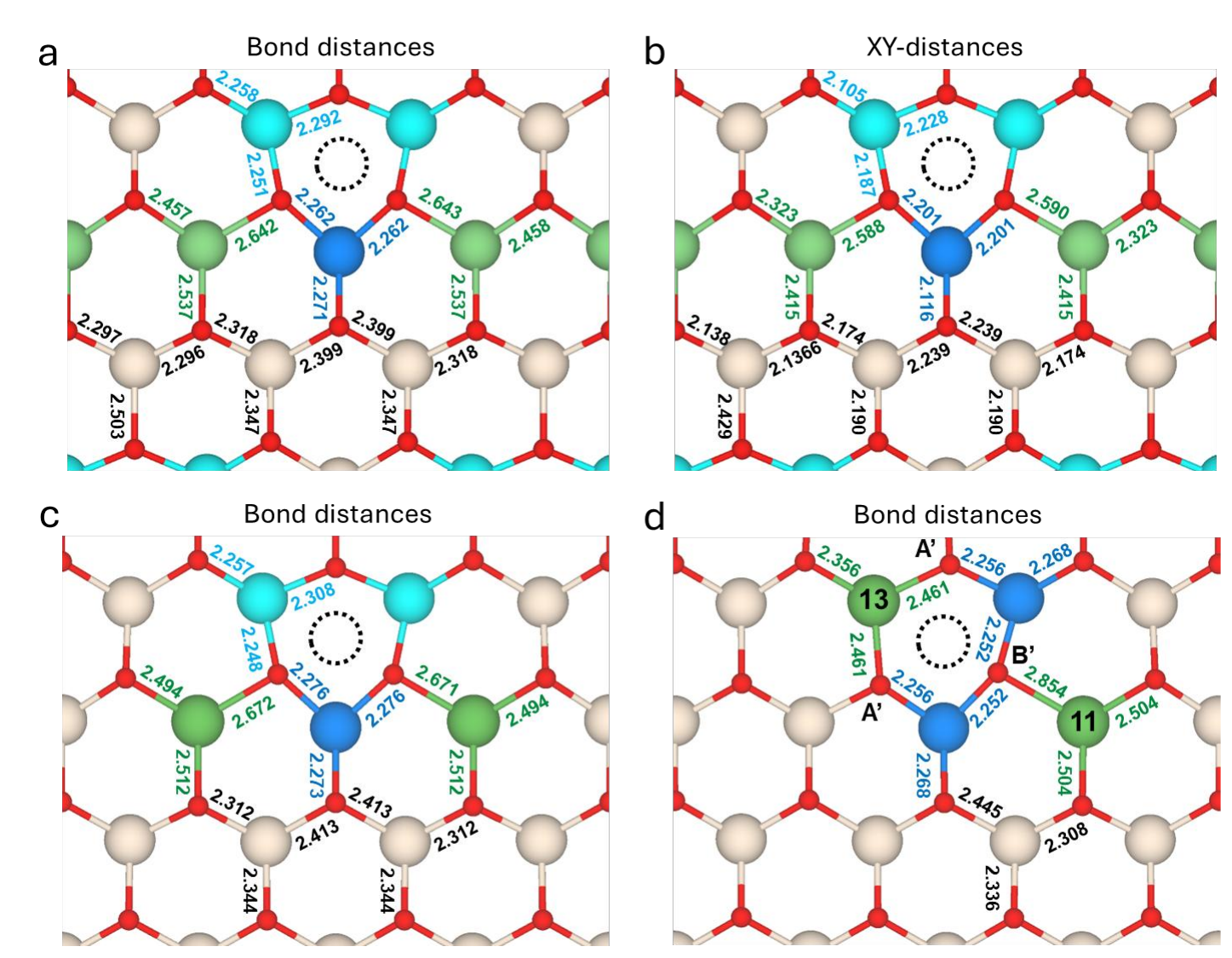}
\caption{{\bf Surface atomic relaxations near a subsurface oxygen vacancy.}
DFT optimized bond distances for the energetically most stable configuration of the two Ce$^{3+}$ atoms produced by a subsurface oxygen vacancy (SSO$_V$) using a (3$\times$3) slab ({\bf a}) and the two most stable configurations using a (4$\times$4) slab ({\bf c} and {\bf d}).
Corresponding X-Y distances of the (3$\times$3) structure are displayed in {\bf b}.
The bond distances are indicated in \AA.
The bond distance on clean ceria is 2.369 Å. O vacancy marked with a dotted circle, Ce$^{3+}$ atoms in green, Ce$^{4+}$ atoms neighboring the SSO$_V$ in blue (dark blue for the one that additionally neighbors the two Ce$^{3+}$), other Ce$^{4+}$ in sand, surface O in red.
Oxygen atoms neighboring the vacancy are also labeled for reference.
}
\label{FigS9}
\end{figure}